\documentclass[%
 amsmath,amssymb,
 reprint,%
]{revtex4-1}

\usepackage{graphicx}
\usepackage{dcolumn}
\usepackage{bm}

\usepackage[utf8]{inputenc}
\usepackage[T1]{fontenc}
\usepackage{mathptmx}
\usepackage{etoolbox}

\usepackage{gnuplottex}
\usepackage{mathtools}
\usepackage{float}
\usepackage{array}
\usepackage[english]{babel}
\usepackage[utf8]{inputenc}
\usepackage{amsmath}
\usepackage[table]{xcolor}
\usepackage{amsfonts}
\usepackage{graphicx}
\usepackage{dsfont}
\usepackage[colorlinks=true, linkcolor=blue, citecolor=blue]{hyperref}
\usepackage{algorithm}
\usepackage[noend]{algpseudocode}
\usepackage{bbold}
\usepackage{mathtools}
\usepackage{amssymb}
\usepackage{bm}
\usepackage{bbm}
\usepackage{tabularx, booktabs}
\usepackage{xspace}
\usepackage{listings}
\usepackage{tikz}
\usepackage{bbding}
\usepackage{physics}
\usepackage{ulem}

\newcommand{\corr}[1]{\textcolor{black}{#1}}

\makeatletter
\def\@email#1#2{%
 \endgroup
 \patchcmd{\titleblock@produce}
  {\frontmatter@RRAPformat}
  {\frontmatter@RRAPformat{\produce@RRAP{*#1\href{mailto:#2}{#2}}}\frontmatter@RRAPformat}
  {}{}
}%
\makeatother
\begin{document}

\preprint{AIP/123-QED}

\title[Quantum embedding of multi-orbital fragments using the Block-Householder-transformation]{Quantum embedding of multi-orbital fragments using the Block-Householder-transformation} 

\author{Saad Yalouz*}
\email{yalouzsaad@gmail.com}
\affiliation{Laboratoire de Chimie Quantique, Institut de Chimie,
CNRS/Université de Strasbourg, 4 rue Blaise Pascal, 67000 Strasbourg, France}

\author{Sajanthan Sekaran} 
\affiliation{Laboratoire de Chimie Quantique, Institut de Chimie,
CNRS/Université de Strasbourg, 4 rue Blaise Pascal, 67000 Strasbourg, France}
 
 \author{\corr{Emmanuel Fromager}} 
\affiliation{Laboratoire de Chimie Quantique, Institut de Chimie,
CNRS/Université de Strasbourg, 4 rue Blaise Pascal, 67000 Strasbourg, France}

\author{Matthieu Sauban\`ere}
\affiliation{ICGM, Université de Montpellier, CNRS, ENSCM, Montpellier, France}

\date{\today}

\begin{abstract}
 
Recently,  some of the authors introduced the use of the Householder transformation as a simple and intuitive method for the embedding of local molecular fragments
(see  Sekaran \textit{et. al.}, Phys. Rev. B 104, 035121 (2021), and Sekaran \textit{et. al.}, Computation 10, 45 (2022)). 
In this work, we present an extension of this approach to the more general case of multi-orbital fragments using the block version of the Householder transformation applied to the one-body reduced density matrix,  unlocking the applicability to general quantum chemistry/condensed-matter physics Hamiltonians.
 A step by step construction of the Block-Householder transformation is presented. Both physical and numerical interest of the approach are highlighted.
The specific mean-field (non-interacting) case is thoroughly detailed as it is shown that the  
embedding of a given $N$ spin-orbitals fragment leads to the generation of two separated sub-systems: a $2N$ spin-orbitals ``fragment+bath'' cluster that exactly contains $N$ electrons, and a remaining cluster's ``environment'' which is described by so-called core electrons. 
We illustrate the use of this transformation in different cases of embedding {scheme} for practical applications.
We particularly focus on the extension of the previously introduced Local Potential Functional Embedding Theory (LPFET) and Householder-transformed Density Matrix Functional Embedding Theory (Ht-DMFET) to the case of multi-orbital fragments.
These calculations are realized on different types of systems such as model Hamiltonians (Hubbard rings) and \textit{ab initio} molecular systems (hydrogen rings).  
\end{abstract}

\maketitle
\section{Introduction}

Over the last decades, quantum embedding theory  has emerged as a promising strategy to describe the electronic structure of large molecules and extended systems in quantum chemistry and condensed matter physics. Many flavours of approaches have been proposed to model strong electronic correlation \corr{including one-body reduced density matrix based methods~\cite{sekaran2021,sekaran2022local,knizia2012density,knizia2013density,sun2016quantum,wu2019projected,JCTC20_Chan_ab-initio_DMET,wouters2016practical}, Green-function based approaches~\cite{Potthoff-VCA,LichtensteinCDMFT,KotliarCDMFT,Hetler1998PRB,georges1992hubbard,georges1996limitdimension,kotliar2004strongly,held2007electronic,zgid2011DMFTquantum} exact-factorization-based methods~\cite{lacombe2020embedding} and dynamical mean-field theory~\cite{georges1992hubbard} to cite but a few (see also Ref.~\cite{sun2016quantum} and references within for more examples of developments).  }

 In practice, the development of embedding  methods is motivated by one central objective: elaborating new approaches that could reduce the computational costs necessary for accessing properties of very large molecular systems which are, in principle, numerically intractable. 
 To proceed,  one usually chooses to adopt a more convenient picture, called the ``{fragment+bath}'' representation,  which drastically simplifies our vision of the full problem.
 {In this} paradigm, the focus is directed to  a specific local \textit{fragment} that represents a sub-part of interest in a large reference system (\textit{e.g.} a molecule or a lattice). 
 The objective is then to accurately describe how this fragment interacts with the rest of the system which is made~up of a very large number of orbitals. 
For this purpose, one commonly relies on the use of particular embedding transformations.
In practice, the role of these transformations is to create an effective sub-system, the so-called ``bath'', whose goal is to mimic the fragment's surrounding.
This sub-system is usually composed of a finite set of external orbitals with no overlap with the fragment's local orbitals.
The union of both orbital subspaces associated to the \textit{fragment} and the \textit{bath}  forms a \textit{cluster} which can be treated at a very high-level of theory due to its small size (compared to the full system). 
This effective treatment of the interactions within a large system is usually taken as a starting point for the elaboration of more involved embedding schemes.
  
In regard to the introduction provided here, it becomes clear that the embedding transformation used to build the ``fragment+bath'' picture plays a central role. 
In this perspective, several approaches have already been proposed in the literature, with a specific focus on the use of unitary transformations. 
At the single particle level, and in the context of lattice density functional theory, T\"ows and Pastor have shown the existence of a generic class of unitary transformations that can partition the one-body reduced density matrix (1-RDM) into a ``fragment+bath'' picture (for single-orbital fragments). The resulting effective two-orbital cluster (obtained from the 1-RDM) can then be used to approximate local contributions of the interaction energy~\cite{tows2011lattice,tows2012spin,tows2014density}. 
Along that line, one should also mention the use of the Singular Value Decomposition (SVD), also known as the ``\textit{Schmidt decomposition}'' (well-known in quantum information theory~\cite{nielsen2001quantum}).
This transformation has been put in the spotlight within the so-called {Density Matrix Renormalization Group (DMRG)~\cite{white1993density,schollwock2005density} and also in the} Density Matrix Embedding Theory (DMET)~\cite{knizia2013density,knizia2012density,wouters2016practical,wouters2016five} which has been proficiently used to describe zero temperature properties of lattices and \textit{ab~initio} Hamiltonians~\cite{leblanc2015solutions,zheng2016ground,cui2020ground,kawano2020comparative,JCTC20_Chan_ab-initio_DMET,pham2019periodic}.
More recently, this transformation has also been used in time- and temperature-dependent DMET thus giving access to more elaborated properties of many-electron systems~\cite{kretchmer2018real,sun2020finite}.  
On a different note, other types of transformations have also been proposed recently with a direct inspiration from relativistic quantum chemistry methods. 
This is the case of the approach presented in Ref.~\cite{muhlbach2018quantum} which is directly inspired from the relativistic exact two-component method~\cite{kutzelnigg2005quasirelativistic, iliavs2007infinite, saue2011relativistic}. 
This partitioning technique turned out to be of particular interest in the context of a projector-based embedding method allowing to simultaneously accelerate and simplify numerical implementations (see details in Ref.~\cite{muhlbach2018quantum} and associated Refs.~\cite{huzinaga1971theory,hegely2016exact}).

In recent works~\cite{sekaran2021,sekaran2022local}, an embedding approach based on the Householder transformation~\cite{householder_unitary_1958} was proposed by some of the authors for the construction of the "fragment+bath" picture via the 1-RDM. 
When applied to this matrix, the Householder transformation fulfils the T\"ows-Pastor conditions~\cite{tows2011lattice} ensuring a maximal decoupling between  the cluster and its environment. From a matrix  point of view, this resumes to the creation of an (almost) block-diagonal structure for the 1-RDM with two main blocks that can be connected: one for the cluster, and another one for its environment. 
Interestingly, it was demonstrated in Ref.~\cite{sekaran2021} that a full block-diagonalization of the 1-RDM can be achieved in the idempotent case (\textit{i.e.} a non-interacting {or mean-field} system), which allows to obtain optimal bath orbitals for a fragment of interest in a simple and automatic way.
This feature recalling notably the concept of optimal bath orbitals produced by the SVD in  DMET~\cite{knizia2013density,knizia2012density,wouters2016practical,wouters2016five}.

Based on the building of the Householder cluster (from which local properties may be extracted), more involved embedding schemes were developed with a specific focus on the \corr{solution of the Hubbard} model~(see Refs~\cite{sekaran2021,sekaran2022local}). 
However, most of the attention had to be spent on the special case of single-orbital fragment due to the limitation of the original Householder transformation.
Naturally, this restriction motivated the research of a more general method to handle multi-orbital fragment: the Block-Householder transformation. Note that several mentions to this method have already been made in our previous work (see Ref.~\cite{sekaran2021}), but its introduction \corr{was only discussed on the surface there}. 
This paper is an opportunity to present this extension with more practical details.

Thus, the present work can be seen as a practical guide for the block-Householder based embedding method for  multi-orbital fragments. {We will show its applicability in both condensed-matter physics and quantum chemistry}. 
{To this purpose,  the Block-Householder transformation is thoroughly detailed along with its corresponding numerical implementation}.  
{ We provide a detailed recipe of different embedding protocols based on this transformation and illustrate their application by means of illustrative numerical examples.}
{To facilitate the reproductibility of our results we use the open access python package \textit{QuantNBody}~\cite{codeQuantNBody} dedicated to the manipulation of quantum many-body systems  and recently proposed by one of the authors (SY). This package allows a systematic construction and diagonalization of the effective embedding Hamiltonians associated to the cluster space.
It also includes our own numerical implementation of the Block-Householder transformation as a ready to use-function.
This function is the one employed to produce all the results presented in this paper}

The paper is organized as follows. In Sec.~\ref{sec:theory} we will  provide a detailed derivation of the embedding based on the Block-Householder transformation. We will here show that an exact partitioning can be reached at the single particle level when considering a mean-field description of the system providing an appealing starting point for embedding methods. 
Then, in Sec.~\ref{sec:Embedding_calculation} we will illustrate how the Block-Householder approach can be used in the context of two methods we recently developed: Ht-DMFET and LPFET. We will present numerical results obtained in the case of model and \textit{ab initio} electronic structure Hamiltonians. Finally, we will bring some conclusion and perspective in the last Sec.~\ref{sec:conclusion}.

  \section{Block-Householder embedding}\label{sec:theory}

In this section, we present how the block-Householder transformation can be employed to build bath orbitals and thus the associated ``{fragment+bath}'' picture. First of all, we briefly remind the details of the original Householder transformation and then present its extended block version. We will then discuss how to use this transformation in an embedding context to partition a 1-RDM and create bath orbitals for a given fragment. 

\begin{figure*} 
  \centering
  \includegraphics[width=\textwidth]{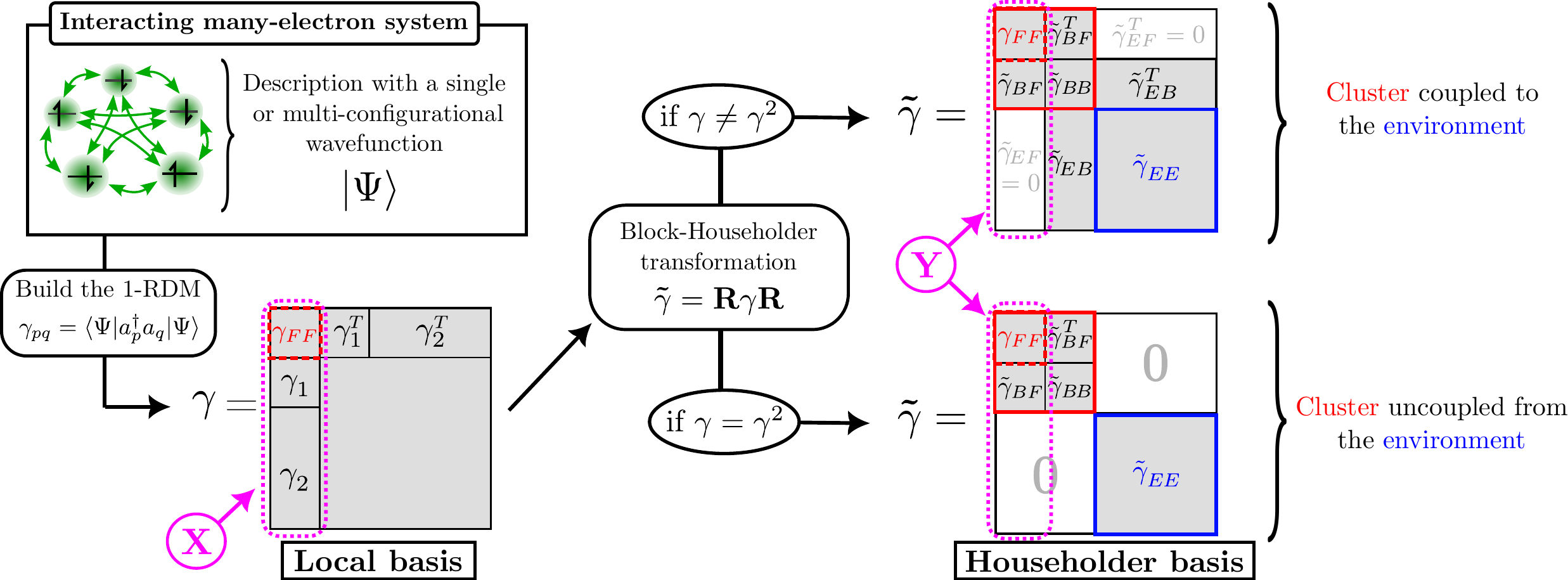}
  \caption{
    \textbf{Steps realized to partition a system's 1-RDM  with the Block-Householder transformation}.
    On the top left, we start with a many-body system described with a wavefunction $\ket{\Psi}$.
    Then, we build the associated 1-RDM and express it in a local orbital basis. In this matrix, we highlight in violet the original column matrix $\bf X$ we want to make sparse.
    For this, we then build the block-Householder transformation ${\bf R}$ and apply it on the 1-RDM such as $\bf  \tilde{ \boldsymbol{\gamma}} = R \boldsymbol{\gamma} R$ (note here that ${\bf R}$ is a functional of the 1-RDM itself). 
    We illustrate then on the right side of the figure the two types of situations we can face.
    On top right, when the 1-RDM is not idempotent  (\textit{i.e.} $ \boldsymbol{\gamma} \neq  \boldsymbol{\gamma}^2$), the partitioning is only partial.
    The cluster is still coupled to its environment as the transformed matrix $\tilde{ \boldsymbol{\gamma}}$ still presents a non-zero block $\tilde{ \boldsymbol{\gamma}}_{EB}$ connecting both subspaces.
    On bottom right, we have the second case where the 1-RDM is idempotent (\textit{i.e.} $ \boldsymbol{\gamma} =  \boldsymbol{\gamma}^2$).
    In this case, the transformed matrix $\tilde{ \boldsymbol{\gamma}}$ becomes strictly block-diagonal (\textit{i.e.} $\tilde{ \boldsymbol{\gamma}}_{EB} = 0$) and the cluster gets totally decoupled from its environment.
    The final sparse matrices $\bf Y = R X$ (originally targeted by the transformation) are highlighted in violet in both transformed 1-RDMs.
  }
  \label{fig:matrix_repressentation}
\end{figure*}

  \subsection{ Householder and Block-Householder transformations }

The original Householder transformation~\cite{householder_unitary_1958} is widely used in computational algebra as a practical tool for the implementation of elaborated matrix manipulations such as QR factorization or tri-diagonalization to cite but a few (see Refs.~\cite{householder_unitary_1958,wilkinson1962householder,martin1968householder}). 
Geometrically, the Householder transformation, noted $\bf R$, represents a reflection which transforms an initial vector $\bf x$ into a final vector $\bf y$ such as $\bf y = \bf R \bf x$. The transformation $\bf R$ is unitary and symmetric ($ \bf R^\dagger R = R R^\dagger = 1$ and $\bf R = R^T$) and can be explicitly written as,
\begin{eqnarray} \label{eq:HH_P_matrix}
 { \bf R( v) } = {\bf 1} - 2 \bf { v  v^{\dagger}} ,
\label{eq:Def_householder_original}
\end{eqnarray}
where $\bf 1$ is the identity matrix and $\bf v$ the normalized Householder vector defined as $\bf v = \frac{\bf x - \bf y}{\abs{\bf x - \bf y}}$, with the dimension $N_{total} \times 1$, and $N_{total}$ the total dimension of the column vector $\bf x$ (\textit{i.e.} its number of elements.)
 In practice, the Householder transformation can be seen as a method for zeroing specific elements of a given column in a reference matrix (which is essential for elaborating more complex transformations). 
 In this context, $\bf x$ represents an initial dense column vector that is transformed into a final vector $\bf y$ which presents the desired sparse structure.


 The block-Householder transformation~\cite{AML99_Rotella_Block_Householder_transf} is an extension of the previous approach that can simultaneously treat (\textit{i.e.} cancel) elements from multiple columns of a given reference rectangular matrix $\bf X$. Similarly to the previous case, the block-Householder transformations, noted $\bf R$, transforms this initial matrix $\bf X$ into a final sparse matrix $\bf Y$, such as $\bf Y =  R  X$. The associated transformation is again unitary and symmetric and takes the following form
\begin{eqnarray}\label{eq:Block_matrix_transformation}
{\bf {R}(V)} =  {\bf1} - 2{\bf V(V^{T}V)^{-1}V^{T}},
\end{eqnarray}
where $\bf V$ has similar dimensions as the matrix we want to transform $\bf X$, namely $N_{total} \times N$. Here, $N_{total}$ is the number of lines of $\bf X$ and $N$ the associated number of columns. In practice, the structure of the $\bf V$ matrix is obtained by the means of involved algebraic manipulations based on the elements of  $\bf X$ which are mathematically cumbersome. For more information about this, we refer the interested reader to the original paper of Rotella and Zambettakis~\cite{AML99_Rotella_Block_Householder_transf} and also to Appendix~\ref{app:block_householder_construction} where we provide details about our implementations of the transformation $\bf R$ (adapted to quantum embedding applications).

  \subsection{ Quantum embedding of a fragment  }

 \subsubsection{ The 1-RDM as a central object }
 
 Following the same philosophy as in our previous work~\cite{sekaran2021}, we will now introduce how to use the Block-Householder transformation to build bath orbitals for a multi-orbital fragment. The main idea is to use this transformation in order to partition the 1-RDM. The protocol presented in the following is summarized and illustrated in Fig.~\ref{fig:matrix_repressentation}. Note here that each spin-orbital basis (delocalized or localized) that we will refer to is supposed to be orthonormalized.
 
The full system under study is described at zero temperature by a quantum state $\ket{\Psi}$ from which we can build the associated 1-RDM, noted  ${   \boldsymbol{\gamma}}$, as follows  
\begin{equation}
     {    \boldsymbol{\gamma}}_{pq} = \bra{ \Psi } \hat{a}^\dagger_p \hat{a}_q  \ket{ \Psi } ,
 \end{equation}
where $\hat{a}^\dagger_p/\hat{a}_p$ represent the electron creation/annihilation operators associated to a given spin-orbital indexed ``$p$''. 
Note here that the 1-RDM is expressed in a local orbital basis (made of $N_{total}$ spin-orbitals in total) as mentioned in the left part of Fig.~\ref{fig:matrix_repressentation}. This choice of basis is necessary to define the spin-orbitals we want to associate to a given reference fragment. This choice has to be made by the user based on some physical criterion (\textit{e.g.} spatial locality of spin-orbitals on a given molecular group).


For sake of simplicity and generality, we will consider here that the upper-left block of the 1-RDM, noted $  \boldsymbol{\gamma}_{FF}$, is associated to the $N$ local spin-orbitals  chosen to belong to a given reference fragment to be embedded.
This square block $  \boldsymbol{\gamma}_{FF}$ of dimension $N \times N$ is represented in Fig.~\ref{fig:matrix_repressentation} with red dashed lines.
Starting from this, the {block-}Householder transformation $\bf R(V)$, and more specifically the matrix $\bold V$, is built from the 1-RDM (\textit{i.e.} the matrix elements underneath~$ \boldsymbol{\gamma}_{FF}$).
By construction, this means that the Householder transformation is itself a functional of the 1-RDM (\textit{i.e.} $\bf R(V) \rightarrow R[ \boldsymbol{\gamma}  ]$). 
In left part of Fig.~\ref{fig:matrix_repressentation}, we represent the 1-RDM elements used for building $\bf R(V)$ as the two-submatrices $ \boldsymbol{\gamma}_1$ and $ \boldsymbol{\gamma}_2$ whose respective dimensions are $ N \times N$ and $(N_{total}-2N) \times N$.
In practice, these two sub-matrices are the ones that compose the initial matrix $\bf X$ we want to transform into the sparse matrix $\bf Y$ as already mentioned in the previous section.
The way the block-Householder transformation is built is motivated by the wish to zero the $ \boldsymbol{\gamma}_2$ block in the final matrix $\bf Y$.
The algebraic manipulations required to built the $\bf R(V)$ unitary matrix are mathematically more involved and cumbersome than in the original Householder method.
For details about our implementations, we invite the interested reader to consult Appendix~\ref{app:block_householder_construction} where we describe all the steps to be implemented  based on an initial a 1-RDM. 

Once the transformation $\bf R(V)$ is constructed numerically from $ \boldsymbol{\gamma}$, one can then build the block-Householder transformed version of the 1-RDM such that
 \begin{equation}
\bf \tilde{ {   \boldsymbol{\gamma}} } =  R(V)   {  \boldsymbol{\gamma}} R(V).
\end{equation}



\corr{The result of this transformation strongly depends on the way we initially describe the system under study:  more precisely, if we use a (multi-configurational) correlated wavefunction such as $\ket{\Psi} = \sum_I C_I \ket{\Phi_I }$,  or a (single configuration) mean-field wavefunction like $\ket{\Psi} = \ket{ \Phi } $.} In the first case, the resulting 1-RDM (and transformed one) will not be idempotent, \textit{i.e.} $ \boldsymbol{\gamma} \neq  \boldsymbol{\gamma}^2$ (and $\tilde{ \boldsymbol{\gamma}}\neq \tilde{ \boldsymbol{\gamma}}^2$), contrarily to the second case, \textit{i.e.} $ \boldsymbol{\gamma} =  \boldsymbol{\gamma}^2$ (and $\tilde{ \boldsymbol{\gamma}} = \tilde{ \boldsymbol{\gamma}}^2$). In the following sub-sections, we will discuss the properties obtained in both situations.

 \subsubsection{ General case :  non-idempotent 1-RDM }
 
The general result obtained after block-Householder transformation of an non-idempotent 1-RDM $ \boldsymbol{\gamma}$ is illustrated in the middle panel of Fig.~\ref{fig:matrix_repressentation}. As shown here, the shape of  $\bf \tilde{ {   \boldsymbol{\gamma}} } =  R(V)   {  \boldsymbol{\gamma}} R(V)$ exhibits two  connected block-diagonal matrices highlighted with full red and blue lines, respectively.

The first main block of dimensions $2N\times 2N$ (delimited with  full red line) represents the \textit{Householder cluster} subspace that encodes how the fragment's local orbitals interact with the rest of the system (\textit{i.e.} the associated bath orbitals). It includes the diagonal fragment and bath sub-blocks, $  \boldsymbol{\gamma}_{FF}$ and $ \tilde{ \boldsymbol{\gamma}}_{BB}$, and the off-diagonal interaction part $\tilde{ \boldsymbol{\gamma}}_{BF}$. Note here the intentional notation $ \boldsymbol{\gamma}_{FF}$ with no ``$\sim$'' to stress that by definition  our implementation of the Block-Householder transformation doesn't change at all the elements from this block (see Appendix~\ref{app:block_householder_construction}). The second main block of dimensions $(N_{total}-2N) \times (N_{total}-2N)$ (delimited with full blue line) encodes the \textit{Householder cluster's environment}. This subspace completes the cluster one and encodes the rest of the orbitals present in the system. 

Note in the present case that these two main diagonal blocks associated to the cluster and its environment are not uncoupled in $\tilde{ \boldsymbol{\gamma}}$. Indeed, as shown in top right part of Fig.~\ref{fig:matrix_repressentation}, there is still the presence of an additional off-diagonal part $\tilde{ \boldsymbol{\gamma}}_{EB}$ that connects them. 
This is a direct consequence of the non-idempotency of the 1-RDM which stems from the multi-configurational character of the original wavefunction $\ket{\Psi} = \sum_I C_I \ket{\Phi_I }$. In this context, the wavefunction encodes ``\textit{charge exchange}'' between orbitals which naturally appears in the 1-RDM. 
While the Householder transformation is built to ensure that we cancel the fragment-environment interaction $\tilde{ \boldsymbol{\gamma}}_{EF}$ (which corresponds to the previous $ \boldsymbol{\gamma}_2$ part in the original local basis shown in Fig.~\ref{fig:matrix_repressentation}), there is nevertheless absolutely no guarantee here that we can also kill the remaining off-diagonal terms connecting the bath to the cluster's environment (\textit{i.e.} $\tilde{ \boldsymbol{\gamma}}_{EB}$ is not a null matrix). 
As a consequence, the partitioning of the orbital space generated by the Block-Householder transformation may lead to fractional number of electrons in both the cluster and the environment which may be problematic for the elaboration of quantum embedding methods.

This ambiguity concerning the number of electrons in each sub-system (\textit{i.e.} the cluster and its environment) can be totally lifted when considering a mono-determinantal (mean-field) description of the system.
In this case, the idempotency of the matrix leads to a series of interesting properties which are introduced in the next sub-section.

 \subsubsection{ Mean-Field case : idempotent 1-RDM }

Interestingly, when applied to a non-interacting (or mean-field) density matrix (\textit{i.e.} built with a single determinant $\ket{\Psi} = \ket{\Phi}$), the partitioning generated by the block-Householder transformation becomes total.
In this case, the off-diagonal terms $\tilde{ \boldsymbol{\gamma}}_{EB}$ naturally cancel leading to a perfect decoupling between the two blocks associated to the cluster and its environment.
The 1-RDM becomes strictly block-diagonal as illustrated in the bottom right part of Fig.~\ref{fig:matrix_repressentation}. 
This feature can be demonstrated based on the idempotent character of the block-Householder transformed 1-RDM (\textit{i.e.} $  \tilde{ \boldsymbol{\gamma}}^2 = \tilde{ \boldsymbol{\gamma}} $) via some matrix manipulation.
For more details about that, we invite the interested reader to consult Appendix~\ref{app:demo_block} of this paper where we provide a  mathematical proof demonstrating this feature. 

Another interesting consequence of this decoupling is the presence of an integer number of electrons in both the cluster and its environment. 
More precisely, we can show that the number of electrons contained in the cluster is exactly equal to the dimension of the $ \boldsymbol{\gamma}_{FF}$ sub-block, namely $N$ (when considering spin-orbitals, or equivalently $2N$ if we refer to spatial orbitals).
In a complementary way, $N_{elec}-N$ electrons are present in the cluster's environment (and respectively $2(N_{elec}-N)$ if we refer to spatial orbitals), with here $N_{elec}$ referring to the total number of electrons in the system.
Here again, cumbersome mathematical developments would be necessary to demonstrate these properties.
For sake of conciseness, we invite the interested reader to see Appendix~\ref{app:integer_number_of_electron} for more details about the associated demonstration.

Having a finite number of electrons in the cluster (and reciprocally in the environment) provides a clear picture of the system's partitioning. Without any ambiguity, we know that no charge fluctuations will arise, and $N$ electrons will always occupy the ``fragment+bath'' subspace. 
This defines an appealing starting point for the development of more elaborated embedding techniques as it becomes, in this case,  more straightforward to employ regular \corr{high-level wave function methods} to describe the electronic correlation within each subspace (\textit{e.g.} configuration interaction). 
    \section{Embedding schemes}\label{sec:Embedding_calculation}

In this section, we use the Block-Householder transformation as a central tool  in the context of two recently introduced embedding schemes: the LPFET and Ht-DMFET methods.
We will detail in the following the philosophy of each scheme and explain how the Block-Householder technique is used in practice. 
Note, however, that the presentation of the methods will be reduced to the essentials, as these have already been largely covered in recent works (see Refs.~\cite{sekaran2021,sekaran2022local}). 
Our main objective here is to illustrate the properties one can expect when considering multi-orbital fragments.


\subsection{ Local Potential Functional Embedding Theory }
 
 \subsubsection{Philosophy of the method} \label{subssubsec:lpfet_philo}
 
{The philosophy of the Local Potential Functional Embedding Theory (LPFET) is to establish a formal connection between DMET~\cite{wouters2016five}(or  \textit{Density Embedding Theory}~\cite{bulik2014density}) and Density Functional Theory (DFT). 
For that purpose, we start from the Hamiltonian $\hat{H}$ of a large complex many-body system we want to solve, and we
 replace it by an effective non-interacting Kohn-Sham Hamiltonian $\hat{H}^{KS}$.
 The latter is composed of the (single-body) Hartree-exchange-correlation potential $\hat{v}^{Hxc}$ supposed to effectively mimic all the electronic many-body effects present in the original system.
 The main idea behind LPFET is then to self-consistently build an approximation of this potential from high-level calculations realized on a reduced-in-size interacting many-body embedding cluster.}
  This paves the way for the development of a DFT where the functional is not fixed {\textit{a priori}}, but rather learned self-consistently from local exploration of the electronic correlation.

To this purpose, one employs the Block-Householder method as an embedding transformation and applies it to the (mean-field) 1-RDM of the original Kohn-Sham system to embed the different fragments of the system. 
For each fragment, the resulting cluster orbitals allow to generate an interacting cluster Hamiltonian one can treat with a high-level wavefunction method (\textit{e.g.} configuration interaction) to capture local electronic correlation.
{The next step is then to adjust the effective Hartree-exchange-correlation potential  $\hat{v}^{Hxc}$ in the Kohn-Sham system to force a matching between the density given by the mean-field 1-RDM and the one obtained with the high-level method in the cluster.
This mapping conditions is inspired from DMET and DET.}
Consequently, the embedding of all the fragments from the non-interacting system will naturally evolve as the auxiliary KS Hamiltonian is progressively adjusted from the local measures of electronic correlation realized in the interacting clusters.
Convergence is reached when the Hartree-exchange-correlation potential stabilizes (\textit{i.e.} when the density from the KS system and the Householder cluster are identical).

{Note that the present description of the LPFET philosophy \corr{has been kept brief}
for sake of conciseness. For more precise details on how the embedding scheme works, we invite the reader to consult the original article~\cite{sekaran2022local}. }


\corr{
As a final note, we want to mention that the LPFET method was originally designed theoretically to tackle the specific case of the homogeneous Hubbard system.
In this context, analytical formulae were derived (see Ref.~\cite{sekaran2022local}) to exactify the connection existing between the chemical potential used in the interacting cluster, and the Kohn-Sham potential employed in the auxiliary mean-field system (which yields bath orbitals). 
As it is, switching to inhomogeneous lattices or \textit{ab initio} molecular systems would require in practice to readapt these developments to build a new basis for an inhomogemous version of LPFET. 
Such developments would represent a specific project in itself leading to a dedicated paper. 
This is clearly outside the scope of the present paper which focuses on the introduction  of the Block-Householder transformation for quantum embedding. We thus leave this for future developments.
}

 \subsubsection{Block Householder in LPFET : application on the 1-D~Hubbard model  }

 We present now some numerical results obtained with the LPFET method using the block-Householder transformation for the embedding of multi-orbital fragments. The original many-body Hamiltonian considered is the uniform \textit{L}=1000 sites 1-D Hubbard lattice (with periodic boundary conditions) whose Hamiltonian reads,
 \begin{equation}
 \begin{split}
     \hat{H} &= -t \sum_{s,\sigma} ( \hat{a}^\dagger_{s,\sigma} \hat{a}_{s+1,\sigma} + \hat{a}^\dagger_{s+1,\sigma} \hat{a}_{s,\sigma}  ) \\
     &+ U \sum_s \hat{n}_{s,\uparrow} n_{s,\downarrow} + \sum_{s,\sigma} v^{ext}_s \hat{n}_{s,\sigma}  ,
 \end{split}
 \label{eq:Hamiltonian_fermi_hubbard}
 \end{equation}
where $t$ is the hopping constant (taken as the energy unit, \textit{i.e.} $t\equiv1$), $U$ is the local on-site repulsion and $\hat{n}_{s,\sigma} = \hat{a}^\dagger_{s,\sigma} \hat{a}_{s,\sigma} $ the number of electron on the local spin-orbital of the site ``$s$'' with spin $\sigma \in \lbrace \uparrow,\downarrow \rbrace $.  

 The auxiliary non-interacting Hamiltonian $\hat{H}^{KS}$ shares a similar shape as the interacting Hamiltonian $\hat{H}$, except that the electron-electron repulsion is here replaced by a local on site Hatree-exchange-correlation potential ${v}^{Hxc}_s$ such that,
 \begin{equation} 
     \hat{H}^{KS} = -t \sum_{s,\sigma} ( \hat{a}^\dagger_{s,\sigma} \hat{a}_{s+1,\sigma} + \hat{a}^\dagger_{s+1,\sigma} \hat{a}_{s,\sigma}  ) +  \sum_{s,\sigma} v_s^{KS}  \hat{n}_{s,\sigma}  ,
 \end{equation}   
 where the Kohn-Sham potential $v_{s}^{KS}$ reads,
 \begin{equation}
 v_s^{KS} = v^{ext}_s + v^{Hxc}_s.
 \end{equation}

\begin{figure}
\centering
\includegraphics[width=\columnwidth]{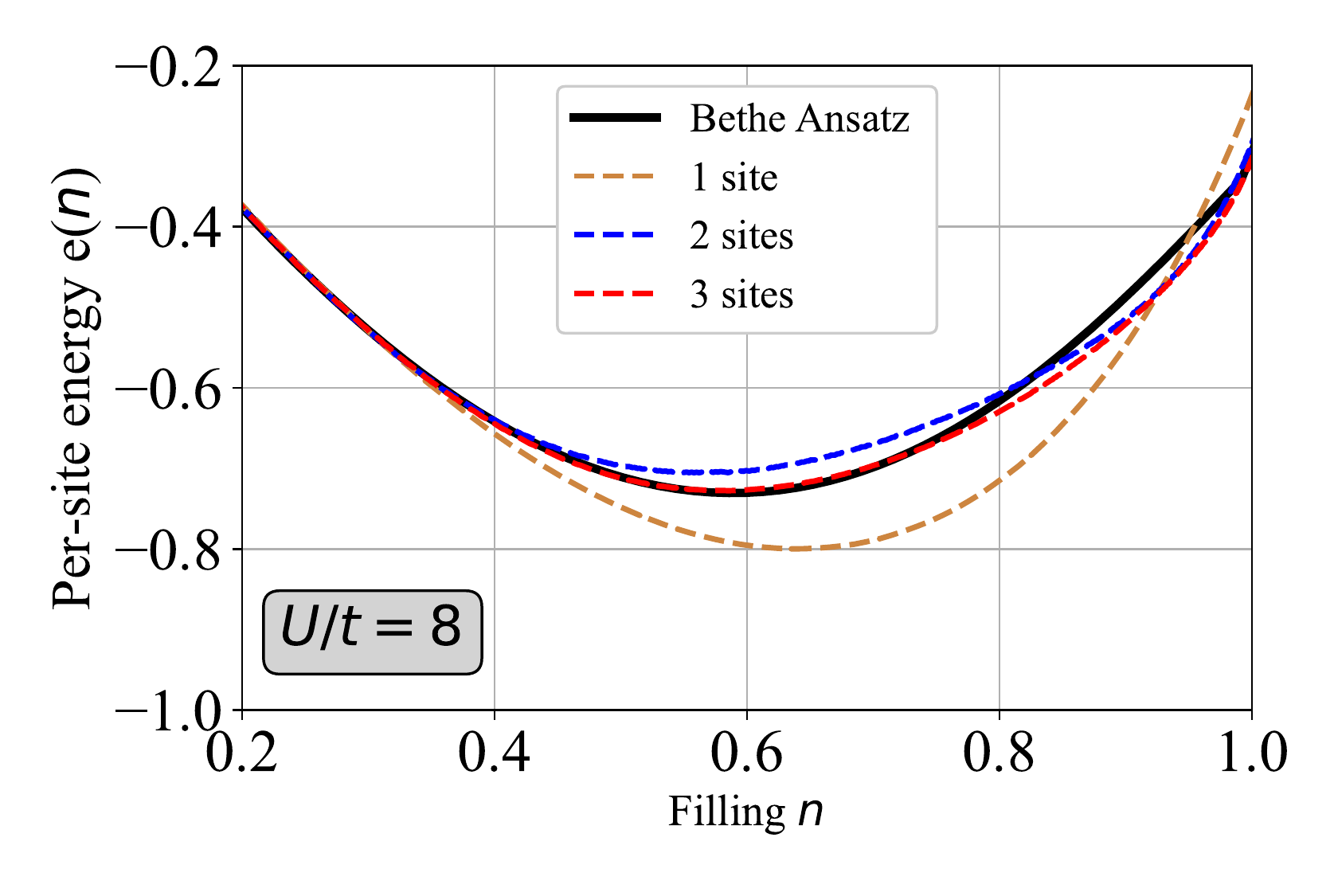}
\caption{
\textbf{LPFET per-site energies as a function of the lattice filling $n$}.  The energy is here plotted  for a strongly correlated case $U$/$t$ = 8. Reference Bethe Ansatz results are shown with a full black line. LPFET results obtained with 1-, 2- and 3-sites fragment are shown with respective colors: orange, blue and red. 
The fragment embedding is realized with the block-Householder transformation technique.
} 
\label{fig:LPFET_persite_energy}
\end{figure}

In Figs.~\ref{fig:LPFET_persite_energy} 
the results obtained for the per-site energy $e(n)$ are presented for three sizes of fragment (one, two and three sites included).
We compare here LPFET results with the exact \textit{Bethe Ansatz} (BA) reference~\cite{lieb1994absence} (black curves) {obtained using the  Knizia's public code~\cite{knizia2012density,shiba1972magnetic}. }

As readily seen in Fig.~\ref{fig:LPFET_persite_energy}, the number of site included in the fragment strongly affects the quality of the per-site energy description for a strongly correlated case (here, $U$/$t$ = 8). Compared to the exact BA curve, the single-site fragment approach (orange curve) produces a poor description of $e(n)$ for $n>0.4$ density regimes and away from half-filling. However, increasing the number of sites in the fragment partitioning allows to improve the description of this quantity. Here, using a three-site fragment partitioning allows to generate a curve that closely follows the exact results along almost all density regimes. A small deviation is observed however when $n>0.8$. Beyond this limit both 2- and 3-sites fragment partitioning lead to a pretty similar evolution.  

\begin{figure}
\centering
\includegraphics[width=\columnwidth]{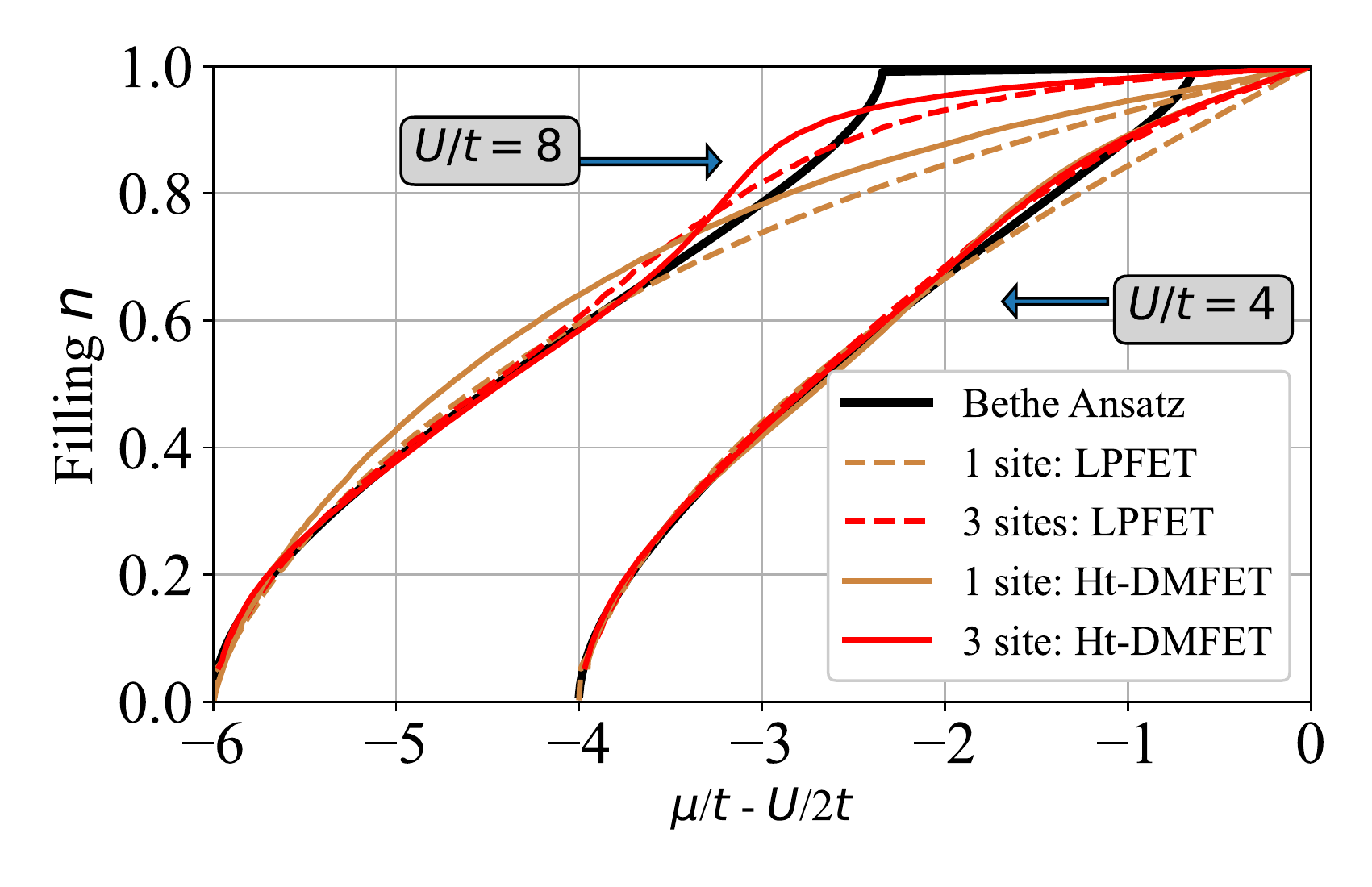}
\caption{ \textbf{LPFET (dashed lines) and Ht-DMFET (full lines) lattice filling $n$ as a function of the lattice chemical potential $\mu$ for a strong correlation regime ($U/t=8$)}. Reference Bethe Ansatz results are shown in full black lines. LPFET and Ht-DMFET results obtained with 1- and 3-sites fragments are shown with respective colors: orange and red. 
The fragment embedding is realized with the block-Householder transformation technique.
Note here that all the results are obtained within the non-interacting bath approximation.}
\label{fig:LPFET_chemical_potential_Hubbard}
\end{figure}

\corr{We now turn toward the density-driven Mott-Hubbard transition which is illustrated in Fig.~\ref{fig:LPFET_chemical_potential_Hubbard} (in dashed lines). 
For sake of comparison, Ht-DMFET results are also presented (in full lines) but will be discussed later. 
As illustrated, whatever the size of the fragment is for the embedding, the LPFET results generated are all very similar and produce an inaccurate description of the gap opening.
For low filling, \textit{i.e.} $n \in [0, 0.6]$, the LPFET curves for 1- and 3-sites fragment globally follow the reference BA evolution.
However, when passing through the Mott-Hubbard transition region which occurs when $n \rightarrow 1$, all LPFET curves deviate. The transition is marked in Fig.~\ref{fig:LPFET_chemical_potential_Hubbard} by the sudden right angle occurring in the BA black curves.
As a result, we see here that the LPFET method fails in reproducing this feature of the system.  Moreover, the quality of the results is not improving when increasing the size of the fragment. This strongly contrasts with the behaviors observed in the case of the per-site energy shown in Fig.~\ref{fig:LPFET_persite_energy}.}



 \subsection{ Householder transformed Density Matrix Functional Embedding Theory  }
 
 \subsubsection{Philosophy of the method } \label{sec:Philo_HtDMFET}

The Householder-transformed Density Matrix Functional Embedding Theory (Ht-DFMET) is an approach that follows closely the ``single-shot'' version of DMET.
Targeting a specific property of a complex interacting many-body system with Hamiltonian $\hat{H}$, the idea here is to replace the expansive diagonalization of this operator by a series of cheaper calculations realized over the different fragments composing the full system. 
In practice, one starts by a mean-field description of the full problem and build the associated idempotent 1-RDM. We express the latter in a local orbital basis to define different fragments in the system. Using an embedding transformation directly on the 1-RDM (\textit{i.e.} the Block-Householder transformation), we then build for each fragment its associated bath orbitals.
The effective Hamiltonians associated to these small-size clusters are then solved with a high-level correlated wavefunction method (\textit{e.g.} configuration interaction). In practice, we observe that the sum of the fragments electronic occupation
may systematically deviate from the one of the original system when solving the interacting problem within the clusters. 
Consequently, in the single-shot approach, we introduce and adjust (self-consistently) a chemical potential $\mu$ on each fragment orbitals within the clusters (in complete analogy with DMET) to match the number of electron in the full system.
Once equality is met, one employs the converged cluster 1- and 2-RDMs to estimate properties of the whole system. 
{Note that while LPFET was exclusively designed for the \textit{non-interacting bath} (NIB) approximation which means that the local electronic repulsion is only present on the fragment orbitals while it is switched off on the bath orbitals, Ht-DMFET can also be employed for the \textit{interacting bath} (IB) approximation which essentially indicates that the on-site electronic repulsion is considered on both the fragment and the bath in the Householder cluster.
More details could be find in Refs.~\cite{wouters2016five,sekaran2021}.}

Naturally this description of the Ht-DMFET philosophy 
\corr{has been kept brief for sake of conciseness}. For more precise details regarding the procedure of the embedding scheme, we invite the reader to consult the original paper~\cite{sekaran2022local}. Note that, in the following, additional details will be provided in due time depending on the type of systems we target (model or \textit{ab initio} one).

 \subsubsection{Block Householder in Ht-DMFET : application on the 1-D~Hubbard model  }

To begin here, we present Ht-DMFET results using the block-Householder transformation to embed multi-orbital fragment in a 1-D Hubbard Hubbard system (whose Hamiltonian is given in Eq.~(\ref{eq:Hamiltonian_fermi_hubbard})). The lattice we consider is the uniform \textit{L}=400 sites with periodic boundary conditions. 
 
\begin{figure}[h]
\centering
\includegraphics[width=\columnwidth]{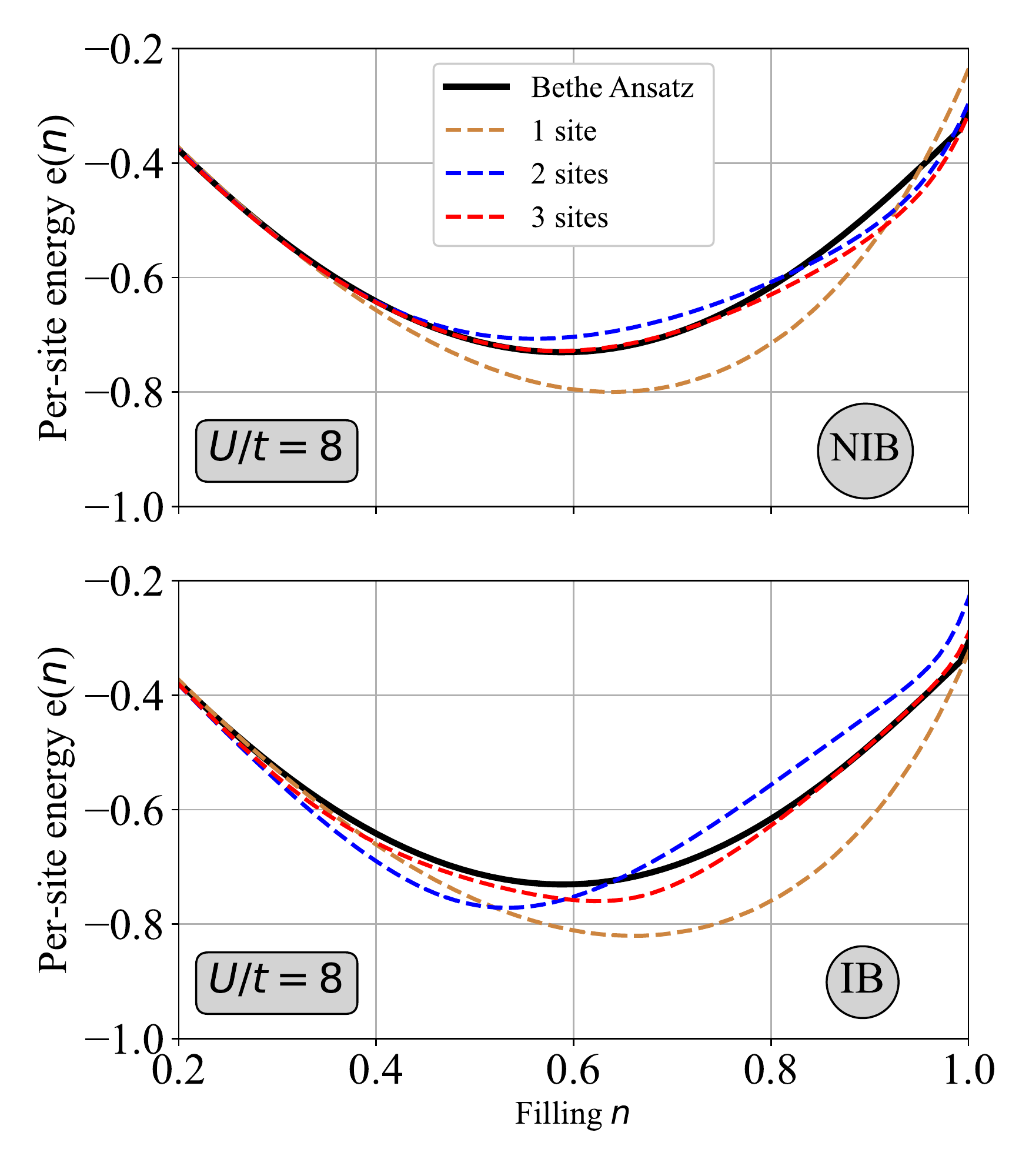}
\caption{
\textbf{Ht-DMFET per-site energies as a function of the lattice filling $n$}.  The energy is here plotted  for a strongly correlated case $U$/$t$ = 8. Reference Bethe Ansatz results are shown with a full black line. Ht-DMFET results obtained with 1-, 2- and 3-sites fragment are shown with respective colors: orange, blue and red. \textbf{Top panel}: Results within the \textit{non-interacting bath} (NIB) approximation. \textbf{Bottom panel}: Results within the \textit{interacting bath} (IB) approximation. The fragment embedding is realized with the block-Householder transformation technique.
}
\label{fig:DMFET_per_site_energy} 
\end{figure}

\begin{figure}
\centering
\includegraphics[width=\columnwidth]{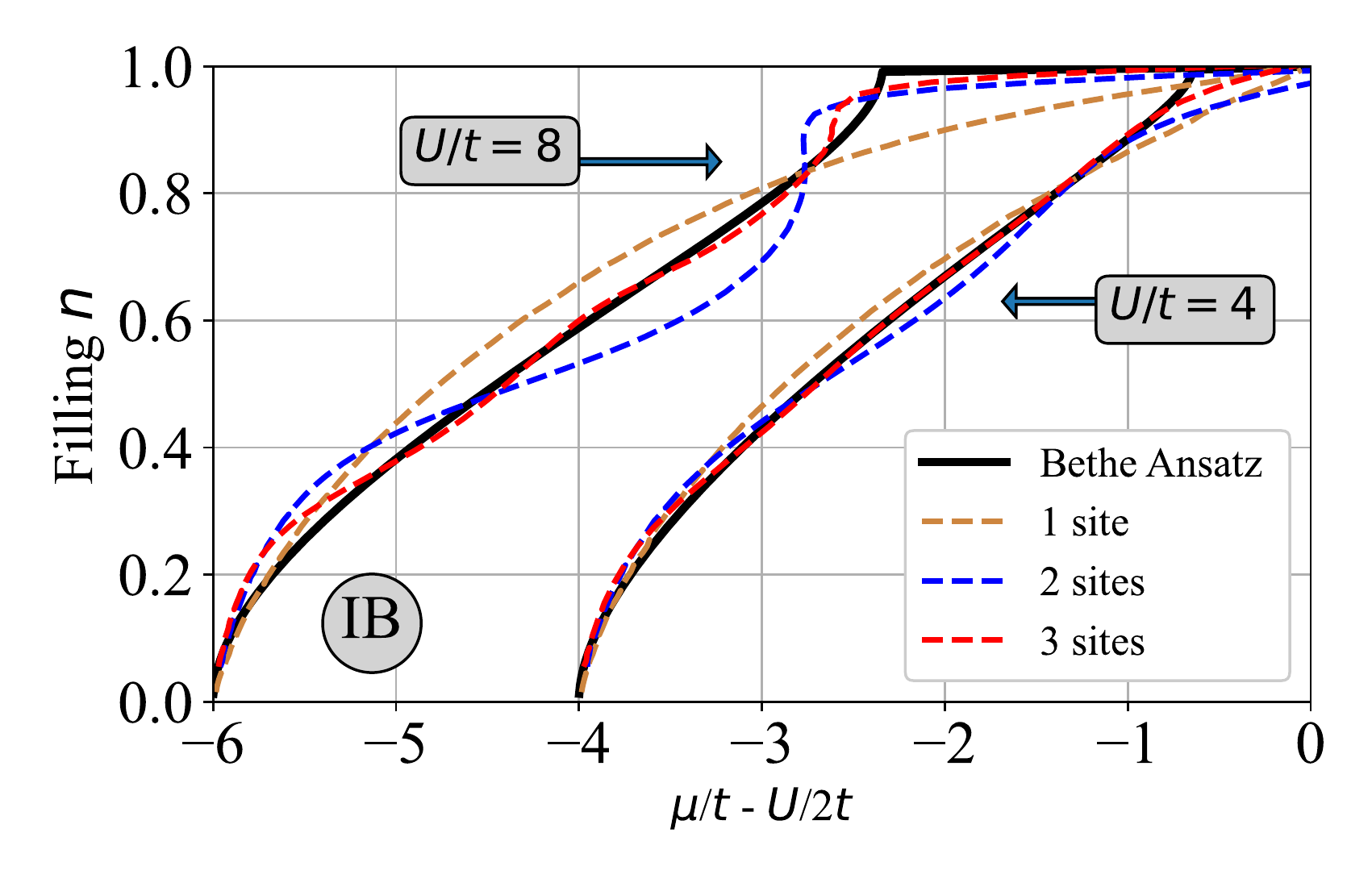}
\caption{ \textbf{Ht-DMFET  lattice filling $n$ as a function of the lattice chemical potential $\mu$}. Here, a strong correlation regime is considered with $U/t=8$. Reference Bethe Ansatz results are shown in full black lines. Ht-DMFET results obtained, within \textit{interacting bath} (IB) approximation, for 1-, 2- and 3-sites fragments are shown with respective colors: orange, blue and red. The fragment embedding is realized with the block-Householder transformation technique.
}
\label{fig:DMFET_chemical_potential_Hubbard} 
\end{figure}


{In Fig.~\ref{fig:DMFET_per_site_energy} we show the evolution of the per-site energy as a function of the lattice filling $n$ for both the NIB and IB approximation}. Focusing first on the NIB scenario (top panel), 
 a careful observer will note the great similarity between Ht-DMFET per-site energy and LPFET results (shown previously in Fig.~\ref{fig:LPFET_persite_energy}). 
 The reason is essentially linked to the fact that in LPFET and in Ht-DMFET (within the non-interacting bath approximation), the occupancy of the fragment orbital is adjusted to be identical to the one of the original interacting system at convergence. 
In the IB approximation of Ht-DMFET (bottom panel), results are less accurate than in the NIB case, especially away from half-filling (\textit{i.e.} when $n \neq 1$).
Nevertheless, here again, the three-site fragment embedding yields the best results compared to the exact BA reference. 
\corr{We now turn toward the density-driven Mott-Hubbard transition as illustrated in Fig.~\ref{fig:LPFET_chemical_potential_Hubbard} and Fig.~\ref{fig:DMFET_chemical_potential_Hubbard} for both the NIB and IB approximation. 
In the NIB case presented in Fig.~\ref{fig:LPFET_chemical_potential_Hubbard}, similar results as LPFET are observed, namely: an inaccurate description of the gap opening. Nevertheless, For 3-sites fragment, it seems that the results are relatively better. Indeed, we observe that the results closely follow the BA reference for $n \in [0, 0.6]$ and the deviation is less pronounced as in LPFET as one can distinguish two changes in the slopes occurring in $n \approx 0.6$ and in $n \approx 0.9$.}
Looking at the Ht-DMFET results in the IB case  (see  Fig.~\ref{fig:DMFET_chemical_potential_Hubbard}), the description of the gap opening phenomenon drastically improves and more specifically when we increase the size of the fragment. 
Note that, these results represent a genuine step forward compared to the ones obtained in our previous work (see Ref.~\cite{sekaran2021}).
Indeed, as shown in our recent paper, when considering the embedding of a single-orbital fragment, no gap opening was observed and this for both the NIB and IB formulations.  
This failure is observable in Fig.~\ref{fig:LPFET_chemical_potential_Hubbard} with the orange curves. 
As shown here, using the Block-Householder transformation allows to go beyond the single-orbital fragment embedding and makes it possible to better describe the sought transition.

\begin{figure*} 
    \centering  
    \includegraphics[width=\textwidth]{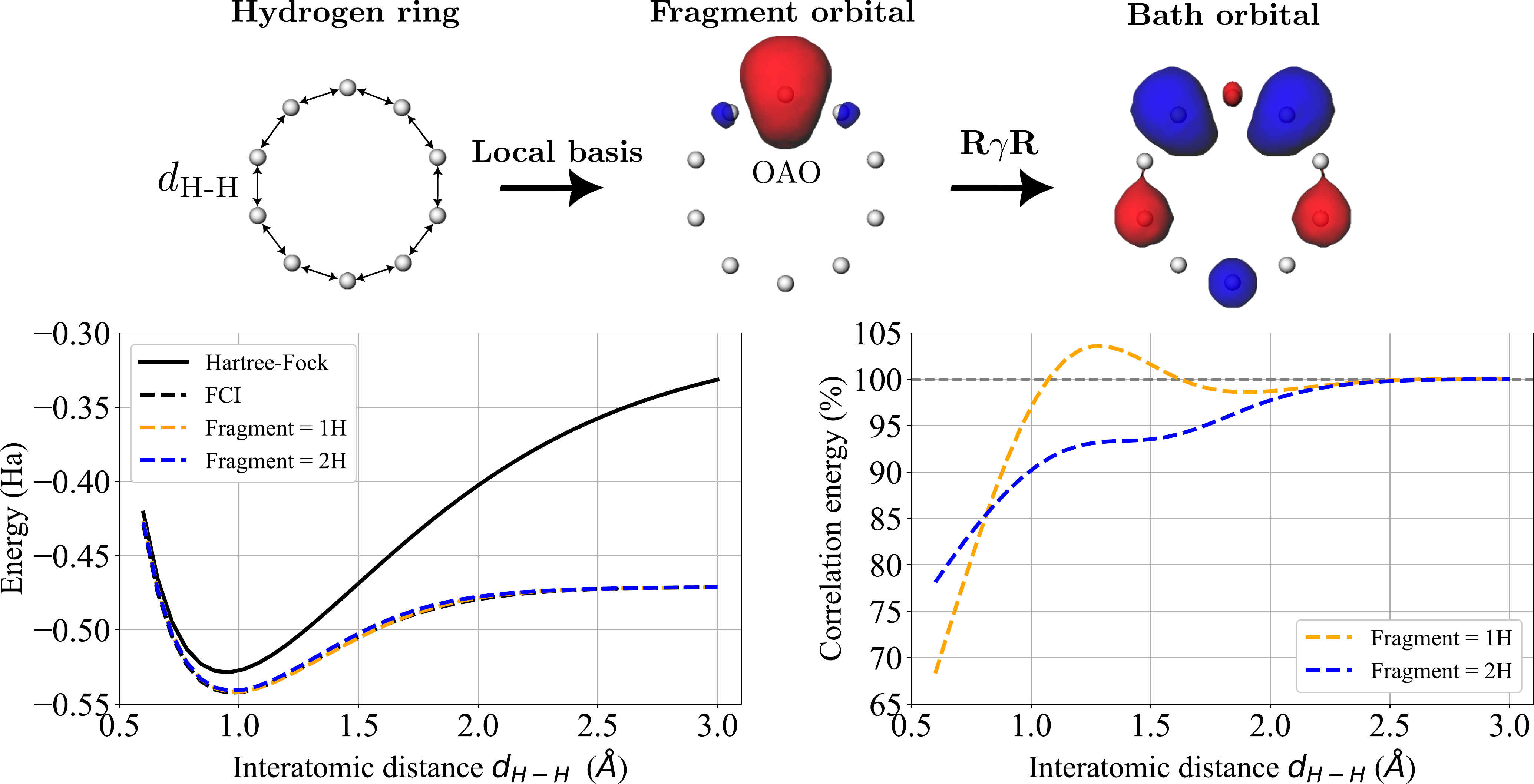} 
    \caption{\textbf{Ht-DMFET calculation on a 10 atoms Hydrogen ring (with STO-6G basis)}. \textbf{Top panel:} In the left plot, we show the hydrogen ring geometry used in our simulation with the interatomic distance $d_\text{H-H}$. On the middle, we illustrate the associated local Orthogonal Atomic Orbital basis used to define molecular fragments. On the right, the resulting shape of a bath orbital is shown for the specific case of a single-orbital fragment (using Block-Householder).  \textbf{Bottom panel:} Comparison of Ht-DMFET energies with FCI (dashed black curve) and Hartree-Fock (full-line black curve). In the left plot, we show the dissociation PES  obtained with the embedding of one and two-atom fragments (orange and blue curves) compared to the exact FCI results (black curve). In the right plot, we show the percentage of correlation energy recovered with both sizes of fragment (this energy is the difference between Hartree-Fock and FCI calculations). }
    \label{fig:H10_Singleshot}
\end{figure*}

\subsubsection{ Block Householder in Ht-DMFET : application on Hydrogen ring H$_{10}$  }



In this last section, we extend the application of the Ht-DMFET using block-Householder embedding transformation to the context of molecular systems. 
To do so, we focus on the realization of embedding calculation  to determine the ground-state energy of an hydrogen ring with 10 atoms. The geometry of the system is shown in the top panel of Fig.~\ref{fig:H10_Singleshot}.

Contrary to the preceding cases where we focused on model Hamiltonians, the realization of a quantum embedding on a chemical system is a bit more involved in practice. 
In this case, the localization of orbitals is an important pre-requisite for the definition of local molecular fragments. 
In practice, many computational methods exist to generate such local basis using an initial mean-field wavefunction (for details, see Ref.~\cite{koridon2021orbital} and references inside).
In the present work, a minimal (STO-6G) basis set is employed to describe the hydrogen orbitals. As a results, one can rely on a more straightforward local orthogonal basis for the embedding: the symmetrically Orthogonalized Atomic Orbitals (OAO) (also know as the Lowdin orbitals see Refs~\cite{lowdin1950non,carlson1957orthogonalization,mayer2002lowdin}). The resulting OAOs are designed to be as close as possible to the original atomic orbitals in a chemical system (in our case the $1s$ orbitals of the hydrogen atoms). 
As illustrated in top panel of Fig.~\ref{fig:H10_Singleshot} (in the middle), the OAOs are indeed naturally localized around their respective atom and these can be straightforwardly used as fragment orbitals. 

To initiate the Ht-DMFET embedding on the hydrogen ring, one first realizes a computationally cheap mean-field calculation (Hartree-Fock) over the whole system to build an idempotent 1-RDM.
We then expressed the associated matrix $ \boldsymbol{\gamma}$ in the local OAO basis and define a given fragment accordingly.
On that basis, we then use the Block-Householder transformation on the 1-RDM to block-diagonalize the latter and build its associated bath orbitals.
An illustration of such bath orbital is given in Fig.~\ref{fig:H10_Singleshot} for the case of a single-orbital fragment. 
As shown here, this bath orbital tends to spatially extends over the rest of the system in a symmetric way with respect to the original OAO chosen. 
Using this approach, we then create a series of clusters calculations that we converge in respect to the Ht-DMFET philosophy (see Sec.~\ref{sec:Philo_HtDMFET}) so that one recovers the total number of electrons in the system. 
After convergence, we build an estimation of the ground-state energy for the full hydrogen system. 
To proceed, we chose to follow the scheme already proposed in DMET for molecular systems: we treat the bath in the so-called interacting picture and we use a democratic partitioning of the local cluster contributions to produce an estimation of the global ground-state energy~(as explained in Ref.~\cite{wouters2016practical,wouters2016five}). 

Following this scheme, we conducted embedding calculations using Block-Householder in Ht-DMFET on the H$_{10}$ hydrogen system. In the bottom panel of Fig.~\ref{fig:H10_Singleshot}, we show the results obtained for the embedding of molecular fragments composed of one and two neighbour Hydrogen atoms (\textit{i.e.} OAOs) with respective orange and blue dashed curves.
For comparison, exact FCI calculations {obtained with the PySCF python package} are also shown (dashed-line black curve) with Hartree-Fock energies (full-line black curve).
The bottom-left panel shows the resulting potential energy surfaces (PES) whereas the bottom-right panel shows the percentage of correlation energies recovered by Ht-DMFET  as  a function of the interatomic distance $d_\text{H-H}$.
As readily seen with the PESs, for both sizes of fragment, the Ht-DMET energies follow the FCI results very closely along the
whole dissociation curve (the FCI curve is almost indistinguishable). 
In similar way to DMET, the evolution of the percentage of correlation energy recovered shows that the Ht-DMFET method is non-variational (as the orange curves goes beyond the $100\%$ of correlation energy). This is expected due to the way the energy is computed here (as explained in Ref.~\cite{wouters2016practical,wouters2016five}).
In practice, no real improvement of the embedding quality is observed when increasing the fragment's size in the Block-Householder embedding. 
This can be related to the intrinsic single shot-nature of Ht-DMFET.

To close this final numerical section, we want to highlight an interesting feature revealed with our results. Here, the set-up chosen for our simulations (namely H$_{10}$ with STO-6G basis and fragments with one and two atoms) is similar to the one used by Wouters \textit{et. al.} to produce single shot DMET calculations as presented in their paper~\cite{wouters2016practical}. The only difference between our work and theirs is the use of the Block-Householder technique to create the ``fragment+bath'' picture instead of the SVD transformation which is central in DMET. Interestingly, we can observe by comparing both works that the curves presented here (see Fig.~\ref{fig:H10_Singleshot}) and in Fig.~3a and b in Ref.~\cite{wouters2016practical} (curves in blue and yellow) seem to be totally equivalent. 
This suggests that the way in which the bath orbitals are produced by the Block-Householder and SVD transformations may be (to some extent) mathematically related.
\corr{ Thus, we tried to address the complex question of the existence of a formal connection for the bath orbitals built with both transformations.  
After some efforts, we finally managed to derive an analytical proof of this feature for the mean-field case (See appendix~\ref{app:proof}).    }

\section{Conclusion}\label{sec:conclusion}

In this paper,  the Block-Householder transformation  was introduced and applied on model and \textit{ab initio} Hamiltonians as an embedding technique for multi-orbital fragments.
In a first part of the paper, we explained how the block-Householder transformation can be used in practice to generate a ``\textit{fragment+bath}'' picture based on the block-diagonalization of the 1-RDM of a given system.
Two particular situations were highlighted.
First, when the 1-RDM is non-idempotent (build with a multi-configurational wavefunction), the block-diagonalization is partial.
This feature is intimately linked to the nature of the wavefunction that describes the system which encodes charge transfer in between orbitals.
As a result, in the transformed 1-RDM, both the cluster and the environment subspace  remain connected which may lead to a non-integer number of electrons.
Second, we demonstrated that when the 1-RDM is idempotent (build with a single determinant), the system subspace decoupling is achieved.
In this case, the block-Householder transformed  1-RDM becomes purely block-diagonal with one block attributed to the cluster and another one to its environment.
In this specific case, the number of electrons present in the cluster is directly proportional to the number $N$ of spin-orbitals we choose to include in the fragment.
This provides a particularly appealing starting point for the implementation of elaborated quantum embedding scheme, which we demonstrated in a second part of this paper with some simulations based on two methods we recently introduced, namely LPFET~\cite{sekaran2022local} and Ht-DMFET~\cite{sekaran2021}. 
In these precedent works, both methods were (essentially) based on the original (and not block) version of the Householder transformation which limited the embedding only to single-orbital fragments. 
Our current work allows then to directly extend these previous developments.
Thus, by applying LPFET to the Hubbard model, we observed that increasing the size of the multi-orbital fragments embedded improves the description of the on-site energy, but not the Mott-Hubbard transition. Similar observation were obtained in the Ht-DMFET scheme within the non-interacting bath approximation. 
Then, turning to the interacting case of Ht-DMFET, a drastic improvement of the gap opening description is observed. 
The quality of the result improving with the size of the embedded fragment. 
This was made possible thanks to the use of the block-version of the Householder transformation.
Finally, we also extended the Ht-DMFET to the quantum chemistry world by using this embedding scheme to study the dissociation of a 10 atoms hydrogen ring. We observed here that the Ht-DMFET results were very accurate compared to the exact FCI simulations.
We also noted that, contrary to the case of the Hubbard model, no clear improvement appeared when increasing the site of the molecular fragment in this context.

Thus,  in this work we demonstrated that the block-Householder transformation is a simple and accessible numerical embedding transformation which can be used in various contexts.
Naturally, all the results presented here motivate a series of questions that could represent interesting starting point for future works.
For example, it would be interesting to investigate how correlated 1-RDM could be used in practice in Ht-DMFET and LPFET. This in order to build bath orbitals which could be more representative of the interacting nature of the original system we want describe.
This would imply to find way to circumvent the problem of non-integer charge in the cluster and thus to re-adapt the block-Householder method to more general case of matrices (and potentially also the embedding method).
Still on block-Householder, another interesting path would be to investigate the use of this method to generate an embedding based on matrices different from the 1-RDM.
To stick to the single-particle level, one could think of the Fock-Matrix, or in a many-body perspective one could try to apply Householder on the 2-RDM, or directly on the density operator $\ketbra{\Psi}{\Psi}$.  
All these ideas are left for future projects and papers.

\section*{Acknowledgements}
E.F. would like to thank Lionel Lacombe for stimulating discussions.
This work was supported by the Interdisciplinary Thematic Institute ITI-CSC via the IdEx Unistra (ANR-10-IDEX-0002) within the program Investissement d’Avenir.   
Financial support was also provided by the ``postdoc researchers'' IdEx 2021 program (University of Strasbourg, grant number: W21RPD04) as well as the DESCARTES (ANR-19-CE29-0002) and CoLab (ANR-19-CE07-0024-02) projects.


\section*{AUTHOR DECLARATIONS}

\subsection*{CONFLICTS OF INTEREST}
The authors have no conflicts to disclose.

\subsection*{AUTHOR CONTRIBUTIONS}
\textbf{Saad Yalouz:} Writing (Lead); Software (equal); Methodology (equal);  Supervision (equal); \textbf{Sajanthan Sekaran:} Writing (supporting); Software (equal); Methodology (equal);
\textbf{Emmanuel Fromager:} Writing (appendix D);
\textbf{Matthieu Saubanère:} Conceptualization (lead); Supervision (equal); Writing (supporting);

\subsection*{ DATA AVAILABILITY }
The data that support the findings of this study are available from the corresponding author upon reasonable request.

\appendix

\numberwithin{equation}{section}
\setcounter{equation}{0}
 
\section{Construction of the Block-Householder transformation matrix} \label{app:block_householder_construction}
 
 In this section, we detail the steps realized to numerically build the block-Householder matrix embedding technique used in our work. The scheme presented here follows closely the work of Rotella and Zambettakis presented in Ref.~\cite{AML99_Rotella_Block_Householder_transf}.

Given a  matrix $\bf V$ of dimension $ N_{total}\times$ $N$ where $N_{total}$ is the total number of local orbitals in a system and $N$ the number of orbital in a reference fragment. We assume here that all the columns of $\bf V $ are are linearly independent, \textit{i.e.} $\mathbf{Rank} ({\bf V}) =  N$ and $N  \leq N_{total}$, then the transformation matrix could be written as follows,
\begin{equation}\label{eq:Block_matrix_transformation_}
{\bf R(V) = 1 -} 2 { \bf V(V^{T}V)^{-1}V^{T} }
\end{equation}
where $\bf 1$ is the identity matrix. For simplicity, $\bf (V^{T}V)^{-1}V^{T}$ which is the Moore-Penrose pseudo-inverse of $\bf V$ will be rewritten as $\bf V^{+}$. Therefore, Eq. (\ref{eq:Block_matrix_transformation_}) becomes
\begin{equation}
{\bf R(V) = 1 }  - 2 { \bf VV^{+}}
\end{equation}
The uniqueness of the pseudo-inverse (Moore-Penrose pseudo-inverse) is only obtained when the four following properties are obeyed:
\begin{equation}
\begin{split}
&P1: \bf VV^{+}V=V \\
&P2: \bf V^{+}VV^{+}=V^{+} \\
&P3: \bf(VV^{+})^{T}=VV^{+} \\
&P4: \bf(V^{+}V)^{T} = V^{+}V
\end{split}
\end{equation}
Let's now consider a block matrix $\bf X$ of dimensions $N_{total} \times N $ with the following form
\begin{equation}\label{submatrix:X}
{\bf X } 
= 
\begin{bmatrix}
 \boldsymbol{\gamma}_{FF}\\
 \boldsymbol{\gamma}_{1} \\
 \boldsymbol{\gamma}_{2}
\end{bmatrix},
\end{equation}
where both $ \boldsymbol{\gamma}_{FF}$ and $ \boldsymbol{\gamma}_{1}$ are of same dimensions $N \times N $ and $ \boldsymbol{\gamma}_{2}$ is of dimension $(N_{total} - 2 \times N) \times N$. Note here that  $ \boldsymbol{\gamma}_{1}$  is assumed to be non-singular and thus invertible. From a physical point of view, this properties can be met when the chosen fragment orbital are not disconnected from its environment. In which case, the embedding would be unnecessary.
Two other matrices of same dimension will be needed in the following, 
\begin{eqnarray}\label{submatrix:X_tilde}
{ \bf \tilde{X}} = 
\begin{bmatrix}
0\\
 \boldsymbol{\gamma}_{1} \\
 \boldsymbol{\gamma}_{2}
\end{bmatrix}
\end{eqnarray}
and
\begin{eqnarray}\label{submatrix:V}
{\bf V} = 
\begin{bmatrix}
0 \\
 \boldsymbol{\gamma}_{1} + {\bf W}\\
 \boldsymbol{\gamma}_{2}
\end{bmatrix}
\end{eqnarray}
We finally obtain,
\begin{eqnarray}\label{eq:R(V)X}
\bf   R(V)X &=& \bf X - 2V(V^{T}V)^{-1}V^{T}X\\
&=& \bf X - 2VV^{+}X
\end{eqnarray}
Some elements of Eq. (\ref{eq:R(V)X}) will be useful later under the following decomposition, 
\begin{equation}
\begin{split}\label{eq:VTV_and_VTX}
\bf V^{T}V &=   ( \boldsymbol{\gamma}_{1} + {\bf W})^{T}( \boldsymbol{\gamma}_{1} + {\bf W}) +  \boldsymbol{\gamma}_{2}^{T} \boldsymbol{\gamma}_{2} \\
&=    \boldsymbol{\gamma}_{1}^{T} \boldsymbol{\gamma}_{1} + {\bf W}^{T} \boldsymbol{\gamma}_{1} +  \boldsymbol{\gamma}_{1}^{T}{\bf W} + {\bf W}^{T}{\bf W} +  \boldsymbol{\gamma}_{2}^{T} \boldsymbol{\gamma}_{2} \\
\bf V^{T}X &= ( \boldsymbol{\gamma}_{1} + {\bf W})^{T} \boldsymbol{\gamma}_{1}  +  \boldsymbol{\gamma}_{2}^{T} \boldsymbol{\gamma}_{2} \\
&=  \boldsymbol{\gamma}_{1}^{T} \boldsymbol{\gamma}_{1} + {\bf W}^{T} \boldsymbol{\gamma}_{1} +  \boldsymbol{\gamma}_{2}^{T} \boldsymbol{\gamma}_{2}
\end{split}
\end{equation}
In the new representation, the transformation of the $\bf X$ block matrix leads to
\begin{eqnarray}
\bf R(V)X = 
\begin{bmatrix} 
 \boldsymbol{\gamma}_{FF} \\
\bf Z_{1} \\
\bf Z_{2}
\end{bmatrix}
= 
\begin{bmatrix}
 \boldsymbol{\gamma}_{FF} \\
\tilde{ \boldsymbol{\gamma}}_{BF} \\
\tilde{ \boldsymbol{\gamma}}_{EF}
\end{bmatrix}
\end{eqnarray}
where the goal of this transformation is to set $\bf Z_{2}$ (\textit{i.e.}~$\tilde{ \boldsymbol{\gamma}}_{EF}$) to 0. Note that one can easily verify thanks to Eqs. (\ref{eq:Block_matrix_transformation}), (\ref{submatrix:V}) and (\ref{eq:R(V)X}), that the first block of the matrix $\bf X$ is unchanged. 

From now on, we'll see how one could choose the appropriate ${\bf W}$ in Eq. (\ref{submatrix:V}) in order to have $\bf Z_{2}=0.$ 
\begin{equation}
\begin{split}
\bf Z_{2} &= \boldsymbol{\gamma}_{2} - 2\boldsymbol{\gamma}_{2}(V^{T}V)^{-1}V^{T}X \\
& = \boldsymbol{\gamma}_{2}(V^{T}V)^{-1} \left[ V^{T}V - 2V^{T}X\right] 
\end{split}
\end{equation}
Therefore
\begin{equation} \label{eq:VTV=2V^TX}
\bf Z_{2} = 0 \Leftrightarrow V^{T}V = 2V^{T}X
\end{equation}
Using both decompositions of Eq.~(\ref{eq:VTV_and_VTX}), we can rewrite Eq.~(\ref{eq:VTV=2V^TX}),
\begin{equation}
\begin{split}
 &  { \bf V^{T}V} =  2 {\bf V^{T}X} \\
\Leftrightarrow &  \boldsymbol{\gamma}_{1}^{T} \boldsymbol{\gamma}_{1} +  {\bf W}^{T} \boldsymbol{\gamma}_{1} +  \boldsymbol{\gamma}_{1}^{T}{\bf W} + {\bf W}^{T}{\bf W} +  \boldsymbol{\gamma}_{2}^{T} \boldsymbol{\gamma}_{2} \\
 = &2 \boldsymbol{\gamma}_{1}^{T} \boldsymbol{\gamma}_{1} + 2{\bf W}^{T} \boldsymbol{\gamma}_{1} + 2 \boldsymbol{\gamma}_{2}^{T} \boldsymbol{\gamma}_{2} \\ 
\end{split}
\label{eq:VTV=2VTX}
\end{equation}
which can be rewritten as,
\begin{equation}\label{eq:further_decomposition}
{\bf W}^{T}{\bf W} +  \boldsymbol{\gamma}_{1}^{T}{\bf W} - {\bf W}^{T} \boldsymbol{\gamma}_{1} =  \boldsymbol{\gamma}_{1}^{T} \boldsymbol{\gamma}_{1} +  \boldsymbol{\gamma}_{2}^{T} \boldsymbol{\gamma}_{2}  
\end{equation}
Once factorized,
\begin{equation}
({\bf W} +  \boldsymbol{\gamma}_{1})^{T}({\bf W} -  \boldsymbol{\gamma}_{1}) =  \boldsymbol{\gamma}_{2}^{T} \boldsymbol{\gamma}_{2}
\end{equation} 
From the property that $\bf A^{T}A$ is symmetric for any matrix $\bf A$, one can deduce that $({\bf W} +  \boldsymbol{\gamma}_{1})^{T}({\bf W} -  \boldsymbol{\gamma}_{1})$ should be symmetric, thus,
\begin{equation}
({\bf W} +  \boldsymbol{\gamma}_{1})^{T}({\bf W} -  \boldsymbol{\gamma}_{1}) = ({\bf W} -  \boldsymbol{\gamma}_{1})^{T}({\bf W} +  \boldsymbol{\gamma}_{1})
\end{equation}
which finally gives,
\begin{equation}\label{eq:gamma1TY_symmetric}
 \boldsymbol{\gamma}_{1}^{T}{\bf W} = {\bf W}^{T} \boldsymbol{\gamma}_{1}
\end{equation}
Finally, we can simplify Eq. (\ref{eq:further_decomposition}) and we obtain the equation that ${\bf W}$ should verify:
\begin{equation}\label{eq:YTY=XTX}
 {\bf W}^{T}{\bf W} =  \boldsymbol{\gamma}_{1}^{T} \boldsymbol{\gamma}_{1} +  \boldsymbol{\gamma}_{2}^{T} \boldsymbol{\gamma}_{2} = \bf \tilde{X}^{T}\tilde{X}
\end{equation}
where the last equality is easily recovered from Eq.~(\ref{submatrix:X_tilde}) and ${\bf W}$ is taken such as $ \boldsymbol{\gamma}_{1}^{T}{\bf W}$ is symmetric (See Eq. \ref{eq:gamma1TY_symmetric}). Rewritting this equation leads to,
\begin{equation}\label{eq:decomposition_YTY=XTX}
 \boldsymbol{\gamma}_{1}^{-T}{\bf W}^{T}{\bf W} \boldsymbol{\gamma}_{1}^{-1} = {\bf 1} +  \boldsymbol{\gamma}_{1}^{-T} \boldsymbol{\gamma}_{2}^{T} \boldsymbol{\gamma}_{2} \boldsymbol{\gamma}_{1}^{-1}
\end{equation}
If one rewrite Eq. (\ref{eq:gamma1TY_symmetric}) into the following form,
\begin{equation}
{\bf W} \boldsymbol{\gamma}_{1}^{-1} =  \boldsymbol{\gamma}_{1}^{-T}{\bf W}^{T}
\end{equation}
and consider $ { \bf M }= {\bf W}{ \boldsymbol{\gamma}_{1}}^{-1}$, Eq. (\ref{eq:decomposition_YTY=XTX}) can be rewritten as
\begin{equation}\label{eq:M-squared}
{\bf M}^{2} = {\bf 1} + \Lambda^{T} \Lambda
\end{equation}
where
\begin{equation} 
\boldsymbol{ \Lambda }=  \boldsymbol{\gamma}_{2} \boldsymbol{\gamma}_{1}^{-1}
\end{equation}
Given that $  \bold{ 1} + \boldsymbol{\Lambda^{T} \Lambda}$ is a positive definite matrix, we look for its square root $\bold{Z}$. Using the Jordan form of  $ \bold{ 1} + \boldsymbol{ \Lambda^{T} \Lambda} $, one can rewrite Eq. (\ref{eq:M-squared}) as,
\begin{equation}\label{eq:M-squared-Jordan's-form}
{\bf 1} + \boldsymbol{ \Lambda^{T} \Lambda} = {\bf P^{T}DP } 
\end{equation}
where $\bf P$ is a unitary matrix and 
\begin{equation}
    {\bf D }= \text{diag}_{i=1}^{N_{total}}\{ d_{i} \}
\end{equation}
with the scalar values $ d_{i} > 0$. Finally, one reads $\bf Z$, the square root of ${\bf M}^{2}$ as,
\begin{eqnarray}\label{eq:Z}
\bf Z = P^{T}\sqrt{D}P
\end{eqnarray}
where ${\bf \sqrt{D}} = \text{diag}_{i=1}^{N_{total}}\{ \sqrt{d_{i}} \}$. Finally, one obtains from $ {\bf M} = {\bf W} \boldsymbol{\gamma}_{1}^{-1}$, Eq. (\ref{eq:M-squared}), (\ref{eq:M-squared-Jordan's-form}) and (\ref{eq:Z}),
\begin{equation}\label{eq:Y_final_form}
{\bf W} = P^{T}\sqrt{D}P \boldsymbol{\gamma}_{1}
\end{equation}
leading to $\bf{Z_{2}} = 0$. Indeed, we know that ${\bf W}$ should verify Eq. (\ref{eq:YTY=XTX}).
\begin{equation}\label{eq:verification_Y}
\begin{split}
{\bf W}^{T}{\bf W} &= ({\bf P^{T}\sqrt{D}P}  \boldsymbol{\gamma}_{1})^{T}({\bf P^{T}\sqrt{D}P} \boldsymbol{\gamma}_{1}) \\
&=  \boldsymbol{\gamma}_{1}^{T} {\bf P^{T}\sqrt{D}PP^{T}\sqrt{D}P }  \boldsymbol{\gamma}_{1} \\
&=  \boldsymbol{\gamma}_{1}^{T} {\bf P^{T}\sqrt{D}\sqrt{D}P}  \boldsymbol{\gamma}_{1} \\
&=  \boldsymbol{\gamma}_{1}^{T} {\bf P^{T}DP}  \boldsymbol{\gamma}_{1} \\
&=  \boldsymbol{\gamma}_{1}^{T}({\bf 1} + \boldsymbol{\Lambda^{T} \Lambda}) \boldsymbol{\gamma}_{1} \\
&=  \boldsymbol{\gamma}_{1}^{T} \boldsymbol{\gamma}_{1} +  \boldsymbol{\gamma}_{1}^{T}( \boldsymbol{\gamma}_{2} \boldsymbol{\gamma}_{1}^{-1})^{T}( \boldsymbol{\gamma}_{2} \boldsymbol{\gamma}_{1}^{-1}) \boldsymbol{\gamma}_{1} \\
&=  \boldsymbol{\gamma}_{1}^{T} \boldsymbol{\gamma}_{1} +  \boldsymbol{\gamma}_{1}^{T} \boldsymbol{\gamma}_{1}^{-T} \boldsymbol{\gamma}_{2}^{T} \boldsymbol{\gamma}_{2} \boldsymbol{\gamma}_{1}^{-1} \boldsymbol{\gamma}_{1} \\
&=  \boldsymbol{\gamma}_{1}^{T} \boldsymbol{\gamma}_{1} +  \boldsymbol{\gamma}_{2}^{T} \boldsymbol{\gamma}_{2} \\
&= \bf \tilde{X}^{T}\tilde{X}
\end{split}
\end{equation}




\section{Block-diagonal shape for idempotent 1-RDM  after block-Householder transformation  }\label{app:demo_block}

In this section, we demonstrate that the block-Householder transformed 1-RDM 
\begin{equation}
  \bf  \tilde{ \boldsymbol{\gamma}} = R(V)  \boldsymbol{\gamma} R(V)
\end{equation}
presents a pure block-diagonal form when starting with an idempotent matrix $ \boldsymbol{\gamma} =  \boldsymbol{\gamma}^2$. The two resulting blocks representing the Householder cluster and its complementary environment.  First, we recall that the block-Householder transformation matrix $\bf R(V)$ is symmetric and unitary, which means  $\bf R(V)=\bf R(V)^{T}$ and $\bf R(V) R(V)^T = R(V)^T R(V) = \mathbf{1}$. As a result, one can show that the original idempotency of $ \boldsymbol{\gamma}$ is a property that is communicated to the transformed 1-RDM as shown below 
\begin{equation}
\begin{split}
\tilde{ \boldsymbol{\gamma}}^{2} &= \bf R(V) \boldsymbol{\gamma} \bf R(V)\bf R(V) \boldsymbol{\gamma} \bf R(V) \\
&= \bf R(V) \boldsymbol{\gamma}^{2} \bf R(V) \\
&= \bf R(V) \boldsymbol{\gamma} \bf R(V) \\
&= \tilde{ \boldsymbol{\gamma}}
\end{split}
\end{equation}
This idempotency can be used to define a set of equalities between  the different blocks of the matrices $ \boldsymbol{\gamma}$ and $ \boldsymbol{\gamma}^{2}$ which are
\begin{equation}\label{eq:matrix_elements_gamma_squared}
\begin{split}
 \boldsymbol{\gamma}_{FF} &=  \boldsymbol{\gamma}_{FF}^{2} +  \tilde{ \boldsymbol{\gamma}}_{BF}^{T}\tilde{ \boldsymbol{\gamma}}_{BF} + \tilde{ \boldsymbol{\gamma}}_{EF}^{T}\tilde{ \boldsymbol{\gamma}}_{EF} \\
 \tilde{ \boldsymbol{\gamma}}_{BF} &= \tilde{ \boldsymbol{\gamma}}_{BF} \boldsymbol{\gamma}_{FF} + \tilde{ \boldsymbol{\gamma}}_{BB}\tilde{ \boldsymbol{\gamma}}_{BF} + \tilde{ \boldsymbol{\gamma}}_{EB}^{T}\tilde{ \boldsymbol{\gamma}}_{EF} \\
\tilde{ \boldsymbol{\gamma}}_{EF} &= \tilde{ \boldsymbol{\gamma}}_{EF} \boldsymbol{\gamma}_{FF} + \tilde{ \boldsymbol{\gamma}}_{EB}\tilde{ \boldsymbol{\gamma}}_{BF} + \tilde{ \boldsymbol{\gamma}}_{EE}\tilde{ \boldsymbol{\gamma}}_{EF} \\
\tilde{ \boldsymbol{\gamma}}_{FB} &= \tilde{ \boldsymbol{\gamma}}_{BF}^{T} \\
\tilde{ \boldsymbol{\gamma}}_{FE} &= \tilde{ \boldsymbol{\gamma}}_{EF}^{T} \\
\tilde{ \boldsymbol{\gamma}}_{BB} &= \tilde{ \boldsymbol{\gamma}}_{BF}\tilde{ \boldsymbol{\gamma}}_{BF}^{T} + \tilde{ \boldsymbol{\gamma}}_{BB}^{2} + \tilde{ \boldsymbol{\gamma}}_{EB}^{T}\tilde{ \boldsymbol{\gamma}}_{EB} \\
\tilde{ \boldsymbol{\gamma}}_{EB} &=  \tilde{ \boldsymbol{\gamma}}_{EF}\tilde{ \boldsymbol{\gamma}}_{BF}^{T} + \tilde{ \boldsymbol{\gamma}}_{EB}\tilde{ \boldsymbol{\gamma}}_{BB} + \tilde{ \boldsymbol{\gamma}}_{EE}\tilde{ \boldsymbol{\gamma}}_{EF} \\
\tilde{ \boldsymbol{\gamma}}_{BE} &= \tilde{ \boldsymbol{\gamma}}_{EB}^{T} 
\end{split}
\end{equation}
By construction, we know that the block $\tilde{ \boldsymbol{\gamma}}_{EF} =  0$ in the Householder representation. This means that
\begin{equation}
\begin{split}
\tilde{ \boldsymbol{\gamma}}_{EF}{ \boldsymbol{\gamma}}_{FF} + \tilde{ \boldsymbol{\gamma}}_{EB}\tilde{ \boldsymbol{\gamma}}_{BF} + \tilde{ \boldsymbol{\gamma}}_{EE}\tilde{ \boldsymbol{\gamma}}_{EF} &= 0 \\
\tilde{ \boldsymbol{\gamma}}_{EB}\tilde{ \boldsymbol{\gamma}}_{BF} &= 0
\end{split}
\end{equation}
Knowing that, $\tilde{ \boldsymbol{\gamma}}_{BF}$ is invertible, we end up with
\begin{equation}
\tilde{ \boldsymbol{\gamma}}_{EB}= 0
\end{equation}
This results shows that starting from an idempotent (mean-field) 1-RDM, no off-diagonal terms can connect the cluster to the environment block in $\tilde{ \boldsymbol{\gamma}}$. As a consequence, the transformed 1-RDM  becomes purely block-diagonal with two perfectly disconnected Householder cluster and environment blocks.

\section{Integer number $N$ of electrons within the Householder cluster for idempotent 1-RDM} \label{app:integer_number_of_electron}

In this section, we demonstrate that the number of electrons contained in the Householder cluster is integer for the particular case of an transformed idempotent matrix $\tilde{ \boldsymbol{\gamma}} = \tilde{ \boldsymbol{\gamma}}^2$. To proceed, one will rely on the estimation of matrix ranks.
In practice, the rank describes the number of independent vectors used to built a representation of a given reference matrix. 
As an illustration, for the full 1-RDM $ \boldsymbol{\gamma}$, the rank is
\begin{equation}
    {\bf Rank}( \boldsymbol{\gamma}) =  {\bf Rank}\left( \sum_i^{N_{elec}/2} \ketbra{\phi_i}{\phi_i} \right)  = \frac{N_{elec}}{2},
\end{equation}
as the (alpha or beta spin) 1-RDM  $ \boldsymbol{\gamma}$ is built from $N_{elec}$ occupied spin-orbitals noted $\ket{\phi_i}$  which are independent and orthogonal. Here, the total number of electrons $N_{elec}$ is supposed to be even. As a result, we see here that the effective number of electrons contained in a given orbital subspace can be directly related to the rank of the matrix built from the same orbitals. 

Based on this observation,  to determine the number of electrons contained in the cluster subspace (and in a complementary way, the number of electrons in the cluster's environment), we will here evaluate  the rank of the Householder cluster block matrix. The targeted block is composed of four sub-matrices which makes the evaluation of the rank as follows (as explained in Ref.~\cite{bernstein2009matrix})
\begin{equation} \label{eq:rank_cluster}
\begin{split}
{\bf Rank} 
\begin{pmatrix}
 \boldsymbol{\gamma}_{FF} & \tilde{ \boldsymbol{\gamma}}_{BF}^T\\
\tilde{ \boldsymbol{\gamma}}_{BF} &  \tilde{ \boldsymbol{\gamma}}_{BB}  
\end{pmatrix}   
&= {\bf Rank} ({ \boldsymbol{\gamma}}_{FF}) \\  
&+ {\bf Rank} (\tilde{ \boldsymbol{\gamma}}_{BB} - \tilde{ \boldsymbol{\gamma}}_{BF}{ \boldsymbol{\gamma}}_{FF}^{-1}\tilde{ \boldsymbol{\gamma}}_{BF}^{T})
\end{split}
\end{equation}
In order to evaluate this rank, one needs several ingredients. First, from the second line in Eq.~(\ref{eq:matrix_elements_gamma_squared}) we know that the following equations holds for the $\tilde{ \boldsymbol{\gamma}}_{BF}$ term
\begin{equation}  
\begin{split}
\tilde{ \boldsymbol{\gamma}}_{BF} &= \tilde{ \boldsymbol{\gamma}}_{BF}{ \boldsymbol{\gamma}}_{FF} + \tilde{ \boldsymbol{\gamma}}_{BB}\tilde{ \boldsymbol{\gamma}}_{BF} + \tilde{ \boldsymbol{\gamma}}^{T}_{EB}\tilde{ \boldsymbol{\gamma}}_{EF}  
\end{split}
\end{equation}
which gives after some manipulations (and convoking the fact that $\tilde{ \boldsymbol{\gamma}}_{EB}=0$ in the idempotent case) the following relation 
\begin{equation} \label{eq:BlockHH_gammabf1}
\begin{split} 
\mathbf{1} - { \boldsymbol{\gamma}}_{FF} &= \tilde{ \boldsymbol{\gamma}}_{BF}^{-1}\tilde{ \boldsymbol{\gamma}}_{BB}\tilde{ \boldsymbol{\gamma}}_{BF}.
\end{split}
\end{equation}
In a similar fashion, one knows by transposing the second line of Eq.~(\ref{eq:matrix_elements_gamma_squared}) that
\begin{equation} 
\begin{split} 
\tilde{ \boldsymbol{\gamma}}_{BF}^{T} &= { \boldsymbol{\gamma}}_{FF}\tilde{ \boldsymbol{\gamma}}_{BF}^{T} + \tilde{ \boldsymbol{\gamma}}_{BF}^{T}\tilde{ \boldsymbol{\gamma}}_{BB} + \tilde{ \boldsymbol{\gamma}}_{EF}^{T}\tilde{ \boldsymbol{\gamma}}^{T}_{EB}  
\end{split}
\end{equation}
which yields the following definition for the $\tilde{ \boldsymbol{\gamma}}_{BB}$ sub-block  (convoking here again the fact that $\tilde{ \boldsymbol{\gamma}}_{EB}=0$ in the idempotent case)
\begin{equation}\label{eq:BlockHH_gammabf2} 
\tilde{ \boldsymbol{\gamma}}_{BB} = \mathbf{1} - \tilde{ \boldsymbol{\gamma}}_{BF}^{-T} { \boldsymbol{\gamma}}_{FF}\tilde{ \boldsymbol{\gamma}}_{BF}. 
\end{equation} 
Then, by multiplying (on the right side) the second line in Eq.~(\ref{eq:matrix_elements_gamma_squared})  by $ \boldsymbol{\gamma}_{FF}$,  one obtains
\begin{equation} 
\begin{split}
\tilde{ \boldsymbol{\gamma}}_{BF}{ \boldsymbol{\gamma}}_{FF} &= \tilde{ \boldsymbol{\gamma}}_{BF} { \boldsymbol{\gamma}}_{FF}^{2} + \tilde{ \boldsymbol{\gamma}}_{BB}\tilde{ \boldsymbol{\gamma}}_{BF}{ \boldsymbol{\gamma}}_{FF} + \tilde{ \boldsymbol{\gamma}}^{T}_{EB}\tilde{ \boldsymbol{\gamma}}_{EF}{ \boldsymbol{\gamma}}_{FF}  
\end{split}
\end{equation}
We then inject in this last equation the definition of ${ \boldsymbol{\gamma}}_{FF}^2$ obtained from the right side of the first line in Eq.~(\ref{eq:matrix_elements_gamma_squared}) and we use  $\tilde{ \boldsymbol{\gamma}}_{EB}=0$ and $\tilde{ \boldsymbol{\gamma}}_{EF}=0$. One thus obtains
\begin{equation} 
\begin{split}
\tilde{ \boldsymbol{\gamma}}_{BF}{ \boldsymbol{\gamma}}_{FF}  
  &= \tilde{ \boldsymbol{\gamma}}_{BF}\left( { \boldsymbol{\gamma}}_{FF} -  \tilde{ \boldsymbol{\gamma}}_{BF}^{T} \tilde{ \boldsymbol{\gamma}}_{BF} \right)  + \tilde{ \boldsymbol{\gamma}}_{BB}\tilde{ \boldsymbol{\gamma}}_{BF} { \boldsymbol{\gamma}}_{FF}  
\end{split}
\end{equation}
After manipulating this equation, one can isolate ${ \boldsymbol{\gamma}}_{FF}$ and find its inverse which is
\begin{equation} 
\begin{split} 
{ \boldsymbol{\gamma}}_{FF}^{-1} &= \tilde{ \boldsymbol{\gamma}}_{BF}^{-1} \tilde{ \boldsymbol{\gamma}}_{BF}^{-T}\tilde{ \boldsymbol{\gamma}}_{BF}^{-1}\tilde{ \boldsymbol{\gamma}}_{BB}\tilde{ \boldsymbol{\gamma}}_{BF}
\end{split}
\end{equation}
Based on this last equation, one can reconstruct the last term of Eq.~(\ref{eq:rank_cluster}),
\begin{equation} 
\begin{split}
\tilde{ \boldsymbol{\gamma}}_{BF}{ \boldsymbol{\gamma}}_{FF}^{-1}\tilde{ \boldsymbol{\gamma}}_{BF}^{T} &= \tilde{ \boldsymbol{\gamma}}_{BF} \left( \tilde{ \boldsymbol{\gamma}}_{BF}^{-1}\tilde{ \boldsymbol{\gamma}}_{BF}^{-T}\tilde{ \boldsymbol{\gamma}}_{BF}^{-1}\tilde{ \boldsymbol{\gamma}}_{BB}\tilde{ \boldsymbol{\gamma}}_{BF} \right) 
\tilde{ \boldsymbol{\gamma}}_{BF}^{T} \\
&= \tilde{ \boldsymbol{\gamma}}_{BF}^{-T}\tilde{ \boldsymbol{\gamma}}_{BF}^{-1}\tilde{ \boldsymbol{\gamma}}_{BB}\tilde{ \boldsymbol{\gamma}}_{BF} \tilde{ \boldsymbol{\gamma}}_{BF}^{T} 
\end{split}
\end{equation}
Then, using both Eq.~(\ref{eq:BlockHH_gammabf1}) and Eq.~(\ref{eq:BlockHH_gammabf2}), one ends up with the following relation
\begin{equation} 
\begin{split}
\tilde{ \boldsymbol{\gamma}}_{BF}{ \boldsymbol{\gamma}}_{FF}^{-1}\tilde{ \boldsymbol{\gamma}}_{BF}^{T} &= \tilde{ \boldsymbol{\gamma}}_{BF}^{-T} \left( \mathbf{1} - { \boldsymbol{\gamma}}_{FF} \right) \tilde{ \boldsymbol{\gamma}}_{BF}^{T} \\
&= \mathbf{1} - \tilde{ \boldsymbol{\gamma}}_{BF}^{-T} { \boldsymbol{\gamma}}_{FF} \tilde{ \boldsymbol{\gamma}}_{BF}^{T} \\
&= \tilde{ \boldsymbol{\gamma}}_{BB} 
\end{split}
\end{equation}
Consequently, by injecting this relation into the second term of the original definition of the rank of the cluster matrix Eq.~(\ref{eq:rank_cluster}), one can show that this quantity reduces to the following simpler form
\begin{equation}  
\begin{split}
{\bf Rank} 
\begin{pmatrix}
 \boldsymbol{\gamma}_{FF} & \tilde{ \boldsymbol{\gamma}}_{BF}^T\\
\tilde{ \boldsymbol{\gamma}}_{BF} &  \tilde{ \boldsymbol{\gamma}}_{BB}  
\end{pmatrix}   =  {\bf Rank} ({ \boldsymbol{\gamma}}_{FF}) 
\end{split}
\end{equation}

Therefore, we demonstrate here that the Householder cluster's rank confounds with the rank of the fragment sub-block $ \boldsymbol{\gamma}_{FF}$. Considering that $ \boldsymbol{\gamma}_{FF}$ is invertible, the associated matrix is then by definition full rank ${\bf Rank} ({ \boldsymbol{\gamma}}_{FF}) = {\bf dim} ({ \boldsymbol{\gamma}}_{FF}) = N$. Here, $N$ is the number of local spin-orbitals of the fragment. We then know that the total number of electrons contained in the cluster subspace is exactly $N$ if we focus on spin-orbitals (or equivalently $2N$ if we refer to spatial orbitals). 
  
 \corr{
 \section{Equivalence between Block-Householder and SVD transformations for idempotent density matrices }\label{app:proof}}
 \corr{
Let us consider the following density-matrix functional spin-orbital
subspace:
\begin{equation}
\begin{split}
\mathcal{B}[\gamma]=\left\{\sum_{p\notin F}\gamma_{pf}
\vert p\rangle
\right\}_{f\in F},
\end{split}
\label{eq:bath_subspace_def_BHH}
\end{equation}
where $\vert p\rangle=\hat{a}_p^\dagger\vert
{\rm vac}\rangle$ refers to a localized (lattice) spin-orbital and [see
Eq.~(\ref{submatrix:X_tilde})]
\begin{equation}
\begin{split}
\left\{\gamma_{pf}\right\}_{p\notin F,f\in F}\equiv { \bf \tilde{X}}.
\end{split}
\label{}
\end{equation}
Since, according to Eqs.~(\ref{eq:Block_matrix_transformation_}) and (\ref{eq:VTV=2VTX}),
\begin{equation}
\begin{split}
{\bf \tilde{X}^T}{\bf R(V)}={\bf \tilde{X}^T}-{\bf V^T}
=\left[
\begin{matrix}
0_{FF}&-{\bf W^T}& 0_{FE} 
\end{matrix}\right],
\end{split}
\label{orthogonality_check_Manu}
\end{equation}  
where we used the following equality [see Eq.~(\ref{submatrix:X}) and
(\ref{submatrix:V})],
\begin{equation}
\begin{split}
{\bf \tilde{X}^T}{\bf V}={\bf X^T}{\bf V},
\end{split}
\label{}
\end{equation}
we conclude, by considering the two zero blocks on the right-hand side
of Eq.~(\ref{orthogonality_check_Manu}), that 
\begin{equation}
\begin{split}
\mathcal{B}[\gamma]=\left(F\oplus E\right)^{\perp}=B.
\end{split}
\label{eq:Bgamma_orth_equal_B}
\end{equation}
Therefore, within the Block Householder transformation, the
bath spin-orbital 
subspace $B$ is simply generated by the (non-orthonormal) spin-orbital
basis introduced in
Eq.~(\ref{eq:bath_subspace_def_BHH}).\\
}

 \corr{
We now turn to regular implementations of
DMET~\cite{wouters2016practical} where the full-system density matrix $\gamma$
is evaluated at the non-interacting or mean-field levels of calculation.
It is therefore idempotent, like in the present work, and the one-electron bath subspace (that we denote
$\overline{B}$ in the following) is constructed by performing an SVD of
the overlap matrix between the fragment spin-orbitals $\left\{f\right\}$
and the (fully) occupied spin-orbitals
$\left\{\kappa\right\}$ in $\gamma$ [see Ref.~\onlinecite{wouters2016practical}]: 
\begin{equation}
\begin{split}
\langle\kappa\vert f
\rangle=\sum^{N}_{\overline{\kappa}=1}\mathcal{U}_{\kappa\overline{\kappa}}\sigma_{\overline{\kappa}}\mathcal{V}^\dagger_{\overline{\kappa}f},
\end{split}
\label{eq:SVD_overlap}
\end{equation}    
where $N={\bf dim}(F)$ and $\left\{\sigma_{\overline{\kappa}}\right\}$
are the singular values. On that basis, we can extract a subset of $N$ occupied
orthonormal spin-orbitals,  
\begin{equation}
\begin{split}
\left\{\vert\overline{\kappa}\rangle=\sum^{occ.}_{\kappa}\mathcal{U}_{\kappa\overline{\kappa}}\vert
\kappa\rangle\right\}_{1\leq \overline{\kappa}\leq N},
\end{split}
\label{eq:kappa_bar_def}
\end{equation}
and consider, for convenience, the following alternative orthonormal
basis for the fragment:
\begin{equation}
\begin{split}
F=\left\{
\vert
f_{\overline{\kappa}}\rangle=\sum_{f\in
F}\mathcal{V}_{f\overline{\kappa}}\vert f\rangle
\right\}_{1\leq \overline{\kappa}\leq N}.
\end{split}
\label{eq:f_kappa_bar_def}
\end{equation}
Note that, according to Eq.~(\ref{eq:SVD_overlap}), 
\begin{equation}
\begin{split}
\langle\overline{\kappa}\vert f_{\overline{\kappa}'}
\rangle=\delta_{\overline{\kappa}\overline{\kappa}'}\sigma_{\overline{\kappa}}.
\end{split}
\label{eq:kappa_bar_dot_f_kappa_bar}
\end{equation}
By keeping only, for each $\vert\overline{\kappa}\rangle$, the
components that
are orthogonal to the fragment subspace,  
\begin{equation}
\begin{split}
\vert\overline{\kappa}\rangle\rightarrow\vert\overline{\kappa}^\perp\rangle=\vert\overline{\kappa}\rangle-\left(\sum^N_{\overline{\kappa}'=1}\vert
f_{\overline{\kappa}'}\rangle\langle
f_{\overline{\kappa}'}\vert\right)\vert\overline{\kappa}\rangle,
\end{split}
\label{}
\end{equation}
{\it i.e.}, according to Eq.~(\ref{eq:kappa_bar_dot_f_kappa_bar}),
\begin{equation}
\begin{split}
\vert\overline{\kappa}^\perp\rangle=\vert\overline{\kappa}\rangle-\sigma_{\overline{\kappa}}\vert
f_{\overline{\kappa}}\rangle,
\end{split}
\label{eq:kappa_orth_final_exp}
\end{equation}
and by normalizing,
\begin{equation}
\begin{split}
\vert\overline{\kappa}^\perp\rangle\rightarrow \vert{b}_{\overline{\kappa}}\rangle=\dfrac{\vert\overline{\kappa}^\perp\rangle}{\sqrt{\langle\overline{\kappa}^\perp\vert
\overline{\kappa}^\perp\rangle}},
\end{split}
\label{eq:dmet_bath_normalization}
\end{equation}
where, according to Eqs.~(\ref{eq:kappa_bar_def}), (\ref{eq:f_kappa_bar_def}), (\ref{eq:kappa_bar_dot_f_kappa_bar}), and (\ref{eq:kappa_orth_final_exp}),
\begin{equation}
\begin{split}
\langle\overline{\kappa}^\perp\vert
\overline{\kappa}^\perp\rangle=1+\sigma^2_{\overline{\kappa}}-2\sigma^2_{\overline{\kappa}}=1-\sigma^2_{\overline{\kappa}},
\end{split}
\label{eq:square_norm_kappabar_orth}
\end{equation}
we obtain the mean-field DMET bath: 
\begin{equation}
\begin{split}
\overline{B}=\left\{\vert{b}_{\overline{\kappa}}\rangle\right\}_{1\leq
\overline{\kappa}\leq N}.
\end{split}
\label{eq:bath_dmet}
\end{equation}
Note that, since $\overline{B}\perp F$, the DMET bath spin-orbitals can be
rewritten as follows [see Eqs.~(\ref{eq:kappa_orth_final_exp}) and (\ref{eq:dmet_bath_normalization})], \begin{equation}
\begin{split}
\vert{b}_{\overline{\kappa}}\rangle&=
\left(\sum_{p\notin F}\vert
p\rangle\langle p\vert\right)\vert{b}_{\overline{\kappa}}\rangle
\\
&=\dfrac{\displaystyle
\left(\sum_{p\notin F}\vert
p\rangle\langle p\vert\right)
\vert\overline{\kappa}\rangle}{\sqrt{\langle\overline{\kappa}^\perp\vert
\overline{\kappa}^\perp\rangle}},
\end{split}
\label{eq:dmet_beth_rewritten}
\end{equation}
thus leading, according to Eq.~(\ref{eq:kappa_bar_def}), to the expression of Eq.~(11) in
Ref.~\onlinecite{wouters2016practical}: 
\begin{equation}
\begin{split}
\vert{b}_{\overline{\kappa}}\rangle=\dfrac{\displaystyle\sum_{p\notin
F}\sum^{occ.}_{\kappa}\langle
p\vert\kappa\rangle\,\mathcal{U}_{\kappa\overline{\kappa}}\vert
p\rangle}{\sqrt{1-\sigma^2_{\overline{\kappa}}}}.
\end{split}
\label{}
\end{equation}
}

 \corr{
We can now show that the bath subspaces as constructed in Eqs.~(\ref{eq:Bgamma_orth_equal_B})
and (\ref{eq:bath_dmet}) are the same. For that purpose, we consider
the DMET embedding cluster's environment spin-orbital subspace
$\overline{E}$ which is defined as follows, 
\begin{equation}
\begin{split}
\end{split}
\overline{E}=\left(F\oplus\overline{B}\right)^\perp.
\label{eq:envcluster_dmet}
\end{equation}
As we assumed that $\gamma$ is idempotent (and therefore consists of
fully occupied or unoccupied spin-orbitals), it comes that  
\begin{equation}
\begin{split}
\gamma_{pf}=\sum^{occ.}_{\kappa}\langle p \vert \kappa\rangle \langle
\kappa \vert f\rangle.
\end{split}
\label{}
\end{equation}
Therefore, for any spin-orbital $\vert \overline{e}\rangle$ in
$\overline{E}$ and any fragment spin-orbital $f$, we have [see
Eq.~(\ref{eq:SVD_overlap})]:
\begin{equation}
\begin{split}
\sum_{p\notin F}\gamma_{pf}\langle \overline{e}\vert p \rangle&=
\sum_{p\notin F}\sum^{occ.}_{\kappa}\langle p \vert \kappa\rangle \langle
\kappa \vert f\rangle \langle \overline{e}\vert p \rangle 
\\
&=\sum_{p\notin F}\sum^{occ.}_{\kappa}\sum^{N}_{\overline{\kappa}=1}\langle p \vert \kappa\rangle\mathcal{U}_{\kappa\overline{\kappa}}\sigma_{\overline{\kappa}}\mathcal{V}^\dagger_{\overline{\kappa}f}\langle \overline{e}\vert p \rangle, 
\end{split}
\label{}
\end{equation}
or, equivalently [see Eqs.~(\ref{eq:kappa_bar_def}),
(\ref{eq:dmet_beth_rewritten}), and (\ref{eq:square_norm_kappabar_orth})],
\begin{equation}
\begin{split}
\sum_{p\notin F}\gamma_{pf}\langle \overline{e}\vert p
\rangle
&=
\sum^{N}_{\overline{\kappa}=1}\sigma_{\overline{\kappa}}\mathcal{V}^\dagger_{\overline{\kappa}f}
\sum_{p\notin F}
\langle    \overline{e}\vert p \rangle\langle p \vert
\overline{\kappa}\rangle
\\
&=
\sum^{N}_{\overline{\kappa}=1}
\sigma_{\overline{\kappa}}
\sqrt{1-\sigma^2_{\overline{\kappa}}}
\;\mathcal{V}^\dagger_{\overline{\kappa}f}
\langle
\overline{e}\vert{b}_{\overline{\kappa}} \rangle,
\end{split}
\label{}
\end{equation}
where $\langle
\overline{e}\vert{b}_{\overline{\kappa}} \rangle=0$, according to
Eqs.~(\ref{eq:bath_dmet}) and (\ref{eq:envcluster_dmet}). As a result,
\begin{equation}
\begin{split}
\sum_{p\notin F}\gamma_{pf}\langle \overline{e}\vert p
\rangle=0,
\end{split}
\label{}
\end{equation}
from which we deduce that [see Eq.~(\ref{eq:bath_subspace_def_BHH})] 
\begin{equation}
\begin{split}
\overline{E}=\left(F\oplus\mathcal{B}[\gamma]\right)^\perp.
\end{split}
\label{}
\end{equation}
Therefore, 
\begin{equation}
\begin{split}
\mathcal{B}[\gamma]=\left(F\oplus\overline{E}\right)^\perp=\overline{B},
\end{split}
\label{}
\end{equation}
which implies, according to Eq.~(\ref{eq:Bgamma_orth_equal_B}), that the
Householder and DMET bath spin-orbital subspaces are identical: 
\begin{equation}
\begin{split}
\overline{B}=B.
\end{split}
\label{}
\end{equation}  
}

\bibliography{biblio.bib}

\end{document}